\documentclass[11pt,fleqn,a4paper]{article} %,draft

\usepackage{amsmath,amssymb,amsthm,amsfonts} % ,enumerate,tikz-cd,backref,refcheck
\usepackage[mathscr]{eucal}
\usepackage{graphics, graphicx}

\usepackage{hyperref}
\hypersetup{colorlinks, linkcolor=blue, citecolor=blue, urlcolor=blue}

\allowdisplaybreaks

\newcommand{\p}{\partial}
\newcommand{\sgn}{\mathop{\rm sgn}\nolimits}

\newlength{\mylength}
\newcommand{\solution}{\hspace*{-\mylength}\bullet\quad}
\newcommand{\solutionRedEq}{\hspace*{-\mylength}\circ\quad}

\newtheorem{theorem}{Theorem}%[section]
\newtheorem{lemma}[theorem]{Lemma}
\newtheorem{corollary}[theorem]{Corollary}
\newtheorem{proposition}[theorem]{Proposition}

{\theoremstyle{definition}
	
	\newtheorem{remark}[theorem]{Remark}
}

\newcommand{\todo}[1][\null]{\ensuremath{\clubsuit}}

\newcommand{\noprint}[1]{}
\newcommand{\lsemioplus}{\mathbin{\mbox{$\lefteqn{\hspace{.77ex}\rule{.4pt}{1.2ex}}{\in}$}}}

\begin{document}

\par\noindent {\LARGE\bf
Point symmetries and Lie reductions\\
of Kadomtsev--Petviashvili equation
\par}

\vspace{4mm}\par\noindent{\large
\large Yevhenii Yu. Chapovskyi$^{\dag}$ and Serhii D. Koval$^{\dag\ddag\S}$
}

\vspace{5.5mm}\par\noindent{\it\small
$^\dag$Institute of Mathematics of NAS of Ukraine, 3 Tereshchenkivska Str., 01024 Kyiv, Ukraine
\par}

\vspace{1.5mm}\par\noindent{\it\small
$^\ddag$Department of Mathematics and Statistics, Memorial University of Newfoundland,\\
$\phantom{^\dag}$\,St.\ John's (NL) A1C 5S7, Canada
\par}

\vspace{1.5mm}\par\noindent{\it\small
$^\S$Department of Mathematics and Computing Science,
Saint Mary’s University,\\
$\phantom{^\ddag}$923 Robie Street, Halifax, NS, B3H3C3, Canada
\par}

\vspace{4mm}\par\noindent
E-mails:
e.chapovskyi@imath.kiev.ua,
skoval@mun.ca.
	
\vspace{4mm}\par\noindent\hspace*{10mm}\parbox{140mm}{\small
We compute the point symmetry pseudogroup of the Kadomtsev--Petviashvili equation using both the direct and the algebraic methods.
We also revisit the classical Lie symmetry analysis of this model in detail, concentrating on hidden infinitesimal and hidden point symmetries.
Additionally, we re-examine the symmetry properties of each codimension-one reduced model
and determine their point symmetry pseudogroups.
}\par\vspace{3mm}

\noprint{


35-XX Partial differential equations
  35Kxx Parabolic equations and parabolic systems {For global analysis, analysis on manifolds, see 58J35}
    35K05 Heat equation
    35K10 Second-order parabolic equations
  35Qxx	Partial differential equations of mathematical physics and other areas of application [See also 35J05, 35J10, 35K05, 35L05]
    35Q53  	KdV equations (Korteweg-de Vries equations)
  35Axx General topics
    35A30 Geometric theory, characteristics, transformations [See also 58J70, 58J72]
  35Bxx Qualitative properties of solutions
    35B06 Symmetries, invariants, etc.
17-XX  Nonassociative rings and algebras
  17Bxx Lie algebras and Lie superalgebras
    17B81 Applications of Lie (super)algebras to physics, etc.
}

\noprint{
\todo potential KP equation, see davi1986b

Hidden symmetries and hidden point symmetries?

Khokhlov--Zabolotskaya equation

Stationary KP in Drinfeld--Sokolov hierarchy?
}

\section{Introduction}

The existence of an infinite hierarchy of generalized (or higher) symmetries
is a well-known characteristic of integrable systems of partial differential equations.
Such hierarchies are often constructed using recursion operators, which map generalized symmetries to generalized symmetries,
and are typically associated with underlying Hamiltonian or bi-Hamiltonian structures. 
A classic example is the Korteweg--de Vries equation, whose bi-Hamiltonian structure gives rise to a recursion operator
that generates an infinite hierarchy of commuting generalized symmetries and,
by virtue of the associated Hamiltonian structures, an infinite hierarchy of conservation laws (see, e.g., \cite[Section~5.2, p.~312]{olve1993A} or \cite{fokas1980a,olve1977a}).

At the same time, identifying recursion operators, bi-Hamiltonian structures,
or infinite hierarchies of generalized symmetries is generally a nontrivial and non-algorithmic task.
In contrast, the computation of infinitesimal or point symmetries can be carried out systematically
using well-established algorithms.
However, the relationship between the structure of the Lie symmetry algebra and the integrability of a partial differential equation is less understood.
For example, the Korteweg–de Vries equation is integrable despite having a four-dimensional Lie invariance algebra (see, e.g., \cite[Example~2.44]{olve1993A}), whereas the Euler system has an infinite-dimensional Lie symmetry algebra but is not integrable (see \cite[Example~2.45]{olve1993A}).
These examples illustrate that the size of the Lie symmetry algebra alone does not provide a reliable criterion for integrability.

The use of symmetry analysis in the investigation of integrable nonlinear partial differential equations has a long history,
with important contributions by Winternitz and his collaborators \cite{davi1985a,davi1986a}, among others.
Systematic investigations of generalized symmetries, conservation laws and structures constituted by symmetry-like objects
have been carried out by various authors,
including Popovych and his collaborators  \cite{boyk2024a,kova2024a,malt2024a,vinn2024a,vinn2026a,vanee2014a}.
These approaches have provided valuable tools for studying integrable systems and their reductions.

For other relevant results on application of Lie symmetry analysis in the study of integrability see
\cite{anco1997a,clar1989a,niki1992a,fush1997a}.

In this paper, we investigate point symmetries of the Kadomtsev--Petviashvili equation
\begin{equation}\label{eq:KP}
\left(u_t+\tfrac32uu_x+\tfrac14u_{xxx}\right)_x=\tfrac34\sigma u_{yy},\quad
\sigma=\pm1,
\end{equation}
which is a two-dimensional generalization of the Korteweg--de Vries equation.
Both these equations stand at the origins of modern integrability theory.
Their study has led to groundbreaking developments such as Hirota's bilinear method,
Miura's transform and Sato's hierarchy,
while also revealing deep connections to representation theory and algebraic geometry~\cite{miwa2000A}.

The Lie invariance algebra of~\eqref{eq:KP} was first computed in~\cite{schw1982a},
and its Lie reductions were studied in~\cite{davi1986a}.
Furthermore, representations for the exponents of the vector fields were derived in~\cite[Section~II.B]{davi1986a},
supplemented by heuristically derived discrete point symmetries.
Because these representations are cumbersome and of limited practical use,
we are motivated to apply modern techniques to compute the complete point symmetry pseudogroup
of the Kadomtsev--Petviashvili equation~\eqref{eq:KP}.
To this end, we apply two distinct approaches: the direct method and the algebraic method.
This dual approach not only allows us to cross-verify the correctness of the computations
but also allows us to compare the efficiency of each method when applied to the single system~\eqref{eq:KP}.

After revisiting the classification of one- and two-dimensional subalgebras
of the maximal Lie invariance algebra of the Kadomtsev–Petviashvili equation~\eqref{eq:KP},
we carry out its codimension-one and codimension-two Lie reductions.
Among the nontrivial reduced submodels of~\eqref{eq:KP}, the Boussinesq and differentiated Korteweg--de Vries equations arise.
We revisit the symmetry analysis of these equations in detail:
specifically, we compute their point symmetry pseudogroups using an algebraic method,
study their discrete, induced, and hidden symmetries (both infinitesimal and point),
perform Lie reductions, and construct exact invariant solutions.
Our study explicitly specifies all transformations required to further reduce these submodels
to integrable equations, such as Painlev\'e I, II and IV, which were omitted in previous studies.

It turns out that codimension-two Lie reductions yield no essentially new solutions,
as all such reductions can be framed as two-step reductions via induced symmetries.
Nevertheless, we present the reduced systems and explicitly connect them to the reductions
with respect to one-dimensional subalgebras.

The structure of the paper is as follows.
In Section~\ref{sec:LieInvAlgebra}, we revisit in detail the structure
of the maximal Lie invariance algebra $\mathfrak g$ of the Kadomtsev--Petviashvili equation~\eqref{eq:KP}.
In particular, we identify a chain of megaideals that will constitute an essential tool for computing
the pseudogroup of point symmetries by the megaideal-based version of the algebraic method.

\looseness=-1
Section~\ref{sec:PSG} is devoted to computing the point symmetry pseudogroup $G$ of~\eqref{eq:KP}
using both the direct and algebraic methods separately.
We also study the structure of $G$, in particular, specify its useful decomposition
and classify discrete elements.
In Section~\ref{sec:ContactSyms}, we compute the pseudogroup of contact symmetries $G_c$ of~\eqref{eq:KP},
and thus establish that it coincides with the first prolongation of $G$, $G_c=G_{(1)}$.

After classifying one- and two-dimensional subalgebras of $\mathfrak g$ in Section~\ref{sec:Subalgebras},
we carry out corresponding codimension-one and -two Lie reductions in Sections~\ref{sec:Codim1Reductions} and~\ref{sec:Codim2Reductions}, respectively.
Furthermore, in Sections~\ref{sec:Reduction11} and~\ref{sec:RedSub12},
we rigorously derive and study the point symmetry pseudogroups of the Boussinesq equation
and the differentiated Korteweg–de Vries equation, respectively.
We also carry out Lie reductions of these submodels to establish all substitutions
required to map them to integrable ordinary differential equations.

We summarize the results of the present paper
and outline research perspectives on symmetries of the Kadomtsev--Petviashvili equation~\eqref{eq:KP} in Section~\ref{sec:Conclusion}.
For readers' convenience,
the constructed exact solutions of the equation~\eqref{eq:KP}
are marked by the bullet symbol~$\bullet$\,
and the constructed exact solutions of the reduced systems of~\eqref{eq:KP}
are marked by the circle symbol~$\circ$\,.

\section{Lie invariance algebra}\label{sec:LieInvAlgebra}

The maximal Lie invariance algebra $\mathfrak g$ of the Kadomtsev--Petviashvili equation~\eqref{eq:KP}
is infinite-dimensional and it is spanned by the vector fields
\begin{gather}\label{eq:LieInvAlgebra}
\begin{split}
&\mathcal T(\tau)=\tau\p_t+\tfrac13\Big(\tfrac23\sigma y^2\tau_{tt}+x\tau_t\Big)\p_x+\tfrac23y\tau_t\p_y
+\tfrac23\Big(\tfrac29\sigma y^2\tau_{ttt}-u\tau_t+\tfrac13x\tau_{tt}\Big)\p_u,
\\
&\mathcal X(\chi)=\chi\p_x+\tfrac23\chi_t\p_u,\quad
\mathcal Y(\rho)=y\rho_t\p_x+\tfrac32\sigma\rho\p_y+\tfrac23y\rho_{tt}\p_u,
\end{split}
\end{gather}%$f\to\tau$, $g\to\rho$ and $h\to\chi$
where $\tau$, $\chi$ and $\rho$ 
run through the set of smooth functions of $t$.
The contact invariance algebra~$\mathfrak g_{\rm c}$ of the equation~\eqref{eq:KP}
coincides with the first prolongation of the algebra~$\mathfrak g$, $\mathfrak g_{\rm c}=\mathfrak g_{(1)}$.

The algebra $\mathfrak g$ was first computed in~\cite{schw1982a} using the system of symbolic computations {\sf REDUCE},
and then recomputed in~\cite{davi1985a} using {\sf Maksyma}.
We double-checked these computations using packages {\sf DESOLV}~\cite{carm2000a} and {\sf Jets} \cite{BaranMarvan,marv2009a} for
the computer algebra system {\sf Maple}.

The nonzero commutation relations between vector fields~\eqref{eq:LieInvAlgebra} spanning $\mathfrak g$ are exhausted by
\begin{gather}\label{eq:CommutatorsInG}
\begin{split}
&[\mathcal T(\tau^1),\mathcal T(\tau^2)]=\mathcal T(\tau^1\tau^2_t-\tau^1_t\tau^2),
\\
&[\mathcal T(\tau),\mathcal X(\chi)]=\mathcal X(\tau \chi_t-\tfrac13\tau_t\chi),\quad 
[\mathcal T(\tau),\mathcal Y(\rho)]=\mathcal Y(\tau \rho_t-\tfrac23\tau_t\rho),
\\
&[\mathcal Y(\rho^1),\mathcal Y(\rho^2)]=\tfrac32\sigma\mathcal X(\rho^1\rho^2_t-\rho^1_t\rho^2).
\end{split}
\end{gather}

The algebra $\mathfrak g$ is nonsolvable.
It admits the semidirect sum decomposition given by $\mathfrak g=\mathfrak f\lsemioplus\mathfrak r$,
where $\mathfrak f=\langle\mathcal T(\tau)\rangle$ and $\mathfrak r=\langle\mathcal X(\chi),\mathcal Y(\rho)\rangle$
are a subalgebra and an ideal of~$\mathfrak g$, respectively.
Note that the subalgebra $\mathfrak f$ is isomorphic to the Lie algebra of real vector fields on the real line $\mathbb R$
via the map $\mathcal T(\tau)\mapsto\tau\p_t$ for all $\tau\in C^\infty(\mathbb R)$.

The ideal $\mathfrak r$ is an infinite-dimensional two-step nilpotent Lie algebra.
It is the central extension of two infinite-dimensional abelian Lie algebras,
each of which is isomorphic to $C^\infty(\mathbb R)$ equipped with the trivial bracket.
This abelian extension is determined by a two-cocycle $\omega\colon C^\infty(\mathbb R)\times C^\infty(\mathbb R)\to C^\infty(\mathbb R)$,
defined by $\omega(\tau^1,\tau^2):=\tfrac32\sigma(\tau^1_t\tau^2-\tau^1\tau^2_t)$.

Furthermore, the algebra $\mathfrak r$ naturally inherits the structure of a
left module over the algebra of smooth local functions $C^\infty(\mathbb R)$.
More specifically, given a function $\alpha\in C^\infty(\mathbb R)$,
the left multiplication by it on an arbitrary element of $\mathfrak r$
is defined by $\alpha(\mathcal X(\chi)+\mathcal Y(\rho)):=\mathcal X(\alpha\chi)+\mathcal Y(\alpha\rho)$.
A right module structure can be defined analogously;
so $\mathfrak r$ is a $\alpha\in C^\infty(\mathbb R)$-bimodule.
As a $C^\infty(\mathbb R)$-module, the algebra $\mathfrak r$ is generated by the vector fields
$\mathcal X(1)=\p_x$ and $\mathcal Y(1)=\p_y$.
Under this module action, the algebra~$\mathfrak r$ exhibits the structure of a current algebra $C^\infty(\mathbb R)\otimes_{\mathbb R}\langle\p_x,\p_y\rangle$
over the two-dimensional abelian Lie algebra $\langle\p_x,\p_y\rangle$.
The commutator of two arbitrary elements of this current algebra is given by
\[
[\chi^1\otimes\p_x+\rho^1\otimes\p_y,\chi^2\otimes\p_x+\rho^2\otimes\p_y]:=\tfrac32\sigma(\rho^1_t\rho^2-\rho^1\rho^2_t)\otimes\p_x,
\]
where $\chi^1$, $\chi^2$, $\rho^1$ and $\rho^2$ are smooth functions of $t$.

While the one-parameter transformations corresponding to $\mathcal X(1)$ and $\mathcal Y(1)$ are simply translations in the $x$ and $y$ variables,
the vector field $\mathcal Y(t)$ generates a Galilean boost with a shear in $x$.
Consequently, the one-parameter transformations generated by $\mathcal X(\chi)$ and $\mathcal Y(\rho)$
can be physically interpreted as generalized current shifts in $x$ and current Galilean boosts with shear, respectively.

\begin{lemma}
The radical of~$\mathfrak g$ coincides with the subalgebra $\mathfrak r=\langle\mathcal X(\chi),\mathcal Y(\rho)\rangle$.
\end{lemma}

\begin{proof}
The proof follows the logic of \cite[Lemma~1]{malt2024a}.

From the commutation relations~\eqref{eq:CommutatorsInG}, it follows immediately that $\mathfrak r$ is a solvable ideal of the Lie algebra~$\mathfrak g$.
Since $\mathfrak r^{(2)}=\{0\}$, $\mathfrak r$ is a two-step solvable ideal.
It remains to show that $\mathfrak r$ is a maximal ideal of $\mathfrak g$.

Assume, for the sake of contradiction, that there exists a solvable ideal $\mathfrak s$ of $\mathfrak g$ strictly containing $\mathfrak r$.
This means that $\mathfrak s$ contains an element $\mathcal T(\tau^1)$ for some nonzero smooth function $\tau^1(t)$.
Consider an interval $I\subset\mathbb R$, such that $\tau^1(t)\ne0$ for all $t\in I$;
henceforth, we restrict all parameter functions in $\mathfrak g$ to the interval $I$.
Then, the commutator $[\mathcal T(\tau^1),\mathcal T(\tau)]=\mathcal T(\tau^1_t\tau-\tau^1\tau_t)$ belongs to $\mathfrak s$
for all smooth functions $\tau\in C^\infty(I)$.
By the existence and uniqueness theorem for a linear first-order linear ordinary differential equation,
we have that for any function $\tau\in C^\infty(I)$ there exists $\bar\tau\in C^\infty(I)$ such that $\tau^1_t\tau-\tau^1\tau_t=\bar\tau$.
Moreover, if $\tau$ runs through the space $C^\infty(I)$ then so is $\bar\tau$,
implying that the ideal $\mathfrak s$ contains the subalgebra $\langle\mathcal T(\tau)\rangle$. 
Therefore, $\mathfrak s^{(n)}\supset\langle\mathcal T(\tau)\rangle$ for all $n\in\mathbb N$ as well,
which contradicts the solvability of the ideal $\mathfrak s$.
\end{proof}

To find the complete point-symmetry pseudogroup of the Kadomtsev--Petviashvili equation~\eqref{eq:KP}
using the algebraic method, we should find sufficiently many megaideals of the Lie invariance algebra $\mathfrak g$.
Recall that a {\it megaideal} $\mathfrak m$ of a Lie algebra $\mathfrak g$ 
is a subspace of $\mathfrak g$ that is invariant under any element $\mathfrak T$
of the automorphism group ${\rm Aut}(\mathfrak g)$ of $\mathfrak g$,
$\mathfrak T\mathfrak m\subseteq\mathfrak m$,~\cite{bihlo2011b,popo2003a}.
Another name for megaideal is {\it fully characteristic ideal}, see \cite[Exercise~14.1.1]{hilg2012A}.

It follows from the elementary properties of megaideals
that the radical $\mathfrak m_2:=\mathfrak r$ of the algebra $\mathfrak g$
is its megaideal.
The other obvious megaideals are the derivative of the radical $\mathfrak m_3:=\mathfrak r'$
and the algebra $\mathfrak g$ itself, $\mathfrak m_1:=\mathfrak g$.
Hence, we have the following chain of megaideals
\[
\mathfrak m_1\varsupsetneq\mathfrak m_2\varsupsetneq\mathfrak m_3,
\]
which will be useful in the course of finding the point-symmetry pseudogroup~$G$ of the equation~\eqref{eq:KP}
using the megaideal-based version of Hydon's algebraic method.

\section{Point-symmetry pseudogroup}\label{sec:PSG}

We compute the point-symmetry pseudogroup~$G$ of the Kadomtsev--Petviashvili equation~\eqref{eq:KP}
employing two different approaches, the {\it direct} and the {\it algebraic methods}.
The use of both methods serves a twofold purpose.
First, it provides a robust cross-verification of the computational results through independent pathways.
Second, applying both the methods to a single equation allows us to compare
their respective computational capabilities and efficiency,
which also offers a better understanding of their applicability scope.
While a similar comparative analysis was recently performed for equivalence pseudogroups in~\cite{kova2025a},
such a study has not yet been carried out in the literature specifically for point-symmetry pseudogroups.

\begin{theorem}\label{thm:PointSymGroupKP}
The complete point-symmetry pseudogroup $G$ of the Kadomtsev--Petviashvili equation~\eqref{eq:KP}
is constituted by the transformations of the form
\begin{gather}
\begin{split}\label{eq:PointSymGroup}
\tilde t={}&T(t),\quad
\tilde x=T_t^{1/3}x+\sigma\frac{2T_{tt}}{9T_t^{2/3}}y^2
+\sigma\varepsilon\frac{2Y^0_t}{3T_t^{1/3}}y+X^0(t),\quad 
\tilde y=\varepsilon T_t^{2/3}y+Y^0(t),
\\[0.8ex]
\tilde u={}&\frac1{T_t^{2/3}}u+\frac{2T_{tt}}{9T_{t}^{5/3}}x
+\frac{4\sigma}{9T_t^{1/3}}\left(\frac{T_{tt}}{3T_t^{4/3}} y^2+\varepsilon\frac{Y^0_t}{T_t}y\right)_t
+\frac2{9T_t^2}\big(3 T_tX^0_t-\sigma(Y^0_t)^2\big),
\end{split}
\end{gather}
where $T$, $X^0$ and $Y^0$ are arbitrary smooth functions of $t$ with $T_t\ne0$
and $\varepsilon=\pm1$.
\end{theorem}

Before proceeding to the proof of this theorem,
we discuss the structure of the group $G$,
along with some important corollaries and conventions.

Each element $\Phi$ of the pseudogroup $G$ can be represented as a composition of elementary transformations.
These are point transformations, which are parameterized by a single discrete or a functional parameter:
\[
\arraycolsep=0ex
\begin{array}{lllll}
\mathscr T(T)\colon\    &\tilde t=T,\ \ &\tilde x=T_t^{1/3}x+\sigma\frac{2T_{tt}}{9T_t^{2/3}}y^2,\ \    &\tilde y=T_t^{2/3}y,\ \ \ 
&\tilde u=\frac{1}{T_t^{2/3}}u + \frac{2T_{tt}}{9T_t^{5/3}}x + \frac{4\sigma}{9T_t^{1/3}}\Big(\frac{T_{tt}}{3T_t^{4/3}}\Big)_ty^2,
\\[1.6ex]
\mathscr X(X^0)\colon\  &\tilde t=t,\ \    & \tilde x=x+X^0,\ \                                       &\tilde y=y,\ \
&\tilde u=u+\frac23X_t^0,
\\[1.3ex]
\mathscr Y(Y^0)\colon\  &\tilde t=t,\ \    & \tilde x=x+\frac{2\sigma}3Y_t^0y,\ \                        &\tilde y=y+Y^0,\ \
&\tilde u=u+\frac{4\sigma}9Y_{tt}^0 y-\frac{2\sigma}9(Y_t^0)^2,
\\[1ex]
\mathscr I\colon\       &\tilde t=t,\ \    &\tilde x=x,\ \                                               &\tilde y=-y,\ \ 
&\tilde u=u,
\end{array}
\]
where $T$, $X^0$ and $Y^0$ are arbitrary smooth functions of $t$.
The one-parameter subgroups in the transformation sets $\{\mathscr T(T)\}$, $\{\mathscr X(X^0)\}$ and $\{\mathscr Y(Y^0)\}$
are generated by vector fields $\mathcal T(\tau)$, $\mathcal X(\chi)$ and $\mathcal Y(\rho)$, respectively.
Here the parameter functions run through their specified sets of values.
A transformation $\Phi$ of the form~\eqref{eq:PointSymGroup} can be represented as the composition
\[
\Phi=\mathscr T(T)\circ\mathscr Y(\tilde Y^0)\circ\mathscr X(\tilde X^0)\circ\mathscr I^{\frac{1-\varepsilon}2}
\]
with
\[
\tilde Y^0:=\frac{Y^0}{T_t^{2/3}}\quad\mbox{and}\quad
\tilde X^0:=T_t^{-1/3}\left(X^0-\sigma\frac{2T_{tt}}{3T_t^2}(Y^0)^2\right).
\]

The pseudogroup $G$ can be represented as a semidirect product of 
a pseudosubgroups~$F$ by~$R$, $G=F\ltimes R$, see~\cite[Appendix~A]{kova2024a} for the notions associated with products of pseudogroups.
This factorization is consistent with the decomposition of the Lie algebra $\mathfrak g=\mathfrak f\lsemioplus\mathfrak r$,
which is given in Section~\ref{sec:LieInvAlgebra}.
Here~$R$ and~$F$ are the normal pseudosubgroup of~$G$ and the pseudosubgroup of~$G$
singled out by the constraints $T=t$, $\varepsilon=1$ and $X^0,Y^0=0$, respectively. 
The pseudosubgroup~$F$ is isomorphic to the pseudogroup of local diffeomorphisms of the real line.
The pseudosubgroup~$R$ is constituted by generalized current shifts in $x$,
current Galilean boosts with shear in~$x$
and the involution of alternating the sign of $y$.
That is, $R$ is isomorphic to $\exp(\mathfrak r)\times\mathbb Z_2$,
where $\exp(r)$ is the subgroup generated by point symmetries $\exp(\mathcal X(h))$ and $\exp(\mathcal Y(g))$.
These algebraic properties lead directly to the following assertions.

\begin{corollary}\label{eq:DiscreteSymmetriesOfKP}
(i) The identity component $G_{\rm id}$ of the pseudogroup of point symmetries $G$
of the Kadomtsev--Petviashvili equation~\eqref{eq:KP}
consists of point transformations~\eqref{eq:PointSymGroup} with $T_t>0$ and $\varepsilon=1$.

(ii) A complete list of discrete symmetries of~\eqref{eq:KP} that are independent
up to composing them with each other and with elements of $G_{\rm id}$
are exhausted by two commuting involutions of alternating signs of $(t,x)$
and of $y$,
$\mathscr J\colon (t,x,y,u)\mapsto(-t,-x,y,u)$,
$\mathscr I\colon (t,x,y,u)\mapsto(t,x,-y,u)$.
\end{corollary}

Note that, similarly to \cite[Remark~4]{boyk2024a},
we assume that all the discrete point transformations are globally defined on the entire underlying space
and they absorb all restrictions.
Only under this assumption are the claims regarding the discrete symmetries rigorous.

The quotient group $G/G_{\rm id}$ of the pseudogroup $G$ by its identity component $G_{\rm id}$
is isomorphic to the Klein four-group $\mathbb Z_2\times\mathbb Z_2$.
In fact, the involutions $\mathscr J$ and $\mathscr I$ have been presented in~\cite{davi1985a},
but there was no complete and rigorous proof that there are no other discrete elements of~$G$.

The following two subsections are devoted to proving Theorem~\ref{thm:PointSymGroupKP}
using the direct and the algebraic methods.

\subsection{Direct method}\label{sec:PSGDirectMethod}

The direct method provides the most universal and systematic approach for finding all point symmetries
of a system of differential equations~$\mathcal L$
because it is based on the very definition of a symmetry transformation.
Recall that a point transformation~$\Phi$ acting on the space of independent and dependent variables of $\mathcal L$
is a symmetry of this system if $\Phi$ preserves its solution set.
If, in addition, the system $\mathcal L$ is locally solvable and is of the maximal rank,
then Hadamard's Lemma~\cite[Theorem~2.31]{kunz2015A} implies that a point transformation~$\Phi$
is a symmetry of the system~$\mathcal L$ if and only if it preserves its form (up to a functional multiplier).
This condition allows us to perform the computations as follows.

Consider the most general form of the transformation $\Phi$ of independent and dependent variables with a nonzero Jacobian.
Then express the derivatives of the original dependent variables
with respect to the original independent variables
in terms of the derivatives of the transformed dependent variables 
with respect to the transformed independent variables using the chain rule.
Substituting the derived expressions into the system $\mathcal L$ yields the system $\Phi_*\mathcal L$
which should be identically satisfied on the solution set of the system~$\mathcal L$.
To enforce the last condition, we fix the ranking of the derivatives of
dependent variables consistent with the structure of $\mathcal L$.
We then substitute the expressions for leading derivatives in view of $\mathcal L$
and its differential consequences into $\Phi_*\mathcal L$
and split the resulting system with respect to parametric derivatives of the dependent variables.
For more details as well as applications of the direct method see, e.g., \cite{bihlo2011b,king1998a,kova2023a,kova2023b,kova2024a,opan2017a,popo2008a}.

The Kadomtsev--Petviashvili equation~\eqref{eq:KP} can be transformed into a system in Kovalevskaya form
by introducing the auxiliary dependent variable $\varphi(t,x,y)$ via the relation $u_y=\varphi_x$.
Therefore, the equation~\eqref{eq:KP} is locally solvable and is obviously of the maximal rank,
hence we can apply the above scheme to compute its point-symmetry pseudogroup~$G$.

We expand the equation~\eqref{eq:KP} as
\begin{gather}\label{eq:KPexpanded}
u_{tx}+\tfrac32u_x^2+\tfrac32uu_{xx}+\tfrac14u_{xxxx}=\tfrac34\sigma u_{yy}
\end{gather}
and consider the most general form of the point transformation
acting on the space of the independent and dependent variables $\mathbb R_{t,x,y}^3\times\mathbb R_u$,
\begin{gather*}
\Phi\colon\quad(\tilde t,\tilde x,\tilde y,\tilde u)=(T,X,Y,U)(t,x,y,u)
\end{gather*}
with the nonzero Jacobian, $|\p(T,X,Y,U)/\p(t,x,y,u)|\ne0$.
Using the chain rule, we express all the derivatives of $u$ appearing in~\eqref{eq:KPexpanded}
in terms of the derivatives of $\tilde u$ with respect to $\tilde t$, $\tilde x$ and $\tilde y$,
and substitute
$-\frac32\tilde u_{\tilde x}^2-\frac32\tilde u\tilde u_{\tilde x\tilde x}
-\frac14\tilde u_{\tilde x\tilde x\tilde x\tilde x}+\frac34\sigma\tilde u_{\tilde y\tilde y}$
for $\tilde u_{\tilde t\tilde x}$.
This gives us the invariance condition,
which, after splitting with respect to parametric derivatives of the dependent variables,
results in an overdetermined system of partial differential equations on the components of $\Phi$.

Writing the invariance condition explicitly using the most general form of $\Phi$ is very involved and, actually, is not useful.
Instead of this, we can consider all possible parametric derivatives of $\tilde u$,
starting from those of the highest order and that do not explicitly appear in~\eqref{eq:KPexpanded},
and split the invariance condition with respect to them.

Applying the total derivative $\mathop{{\rm D}_x^4}$ to the equality $\tilde u(\tilde t,\tilde x,\tilde y)=U(t,x,y,u)$,
where $(\tilde t,\tilde x,\tilde y)=(T,X,Y)(t,x,y,u)$,
we obtain that the expression of $u_{xxxx}$ in terms of the derivatives of $\tilde u$ with respect to $\tilde t$, $\tilde x$ and $\tilde y$
contains the summands
\[
\frac{({\rm D}_xT)^4}{U_u-T_u\tilde u_{\tilde t}-X_u\tilde u_{\tilde x}-Y_u\tilde u_{\tilde y}}\tilde u_{\tilde t\tilde t\tilde t\tilde t}\quad
\mbox{and}\quad
\frac{({\rm D}_xY)^4}{U_u-T_u\tilde u_{\tilde t}-X_u\tilde u_{\tilde x}-Y_u\tilde u_{\tilde y}}\tilde u_{\tilde y\tilde y\tilde y\tilde y}.
\] 
This gives the only occurrence of the jets $\tilde u_{\tilde t\tilde t\tilde t\tilde t}$ and $\tilde u_{\tilde y\tilde y\tilde y\tilde y}$
in the invariance condition,
resulting in the determining equations
\begin{gather*}\label{eq:ConditionsOnTandY}
{\rm D}_xT=T_x+T_uu_x=0\quad\mbox{and}\quad
{\rm D}_xY=Y_x+Y_uu_x=0,
\end{gather*}
respectively.
Splitting the obtained equations with respect to $u_x$ gives us $T_x=T_u=0$ and $Y_x=Y_u=0$.

The jet variable $\tilde u_{\tilde x\tilde x\tilde x\tilde x}$ originates from $u_{tx}$ and in $u_{xxxx}$.
Collecting coefficients of summands involving $\tilde u_{\tilde x\tilde x\tilde x\tilde x}$
results in the determining equation
\[
({\rm D}_xX)\big({\rm D}_tT-({\rm D}_xX)^3\big)=0.
\]
If ${\rm D}_xX=0$, then we have $X_x=X_u=0$,
which, in view of the above constraints on $T$ and $Y$ leads to the vanishing of the Jacobian of $\Phi$.
Thus, ${\rm D}_xX\ne0$, leaving us with ${\rm D}_tT-({\rm D}_xX)^3=0$.
Applying the total derivative ${\rm D}_x$ to the last equality,
we obtain ${\rm D}_x^2X=0$,
whose splitting with respect to $u_{xx}$, $u_x$ and $1$
gives us $X_u=0$ and $X_{xx}=0$.
Therefore,
\[
X=X^1(t,y)x+X^0(t,y).
\]
where $X^1$ and $X^0$ are arbitrary smooth functions of $(t,y)$.
Moreover, the equality ${\rm D}_tT-({\rm D}_xX)^3=0$ can be rewritten as
\begin{gather}\label{eq:RelationTX}
T_t=(X^1)^3.
\end{gather}

In view of $|\p(T,X,Y,U)/\p(t,x,y,u)|\ne0$, the conditions $T_u=X_u=Y_u=0$ imply $U_u\ne0$.

The derivative $\tilde u_{\tilde x\tilde x\tilde x}$ comes only from $u_{xxxx}$
with the coefficient $-4({\rm D}_xX)^3U_u^{-1}(U_{uu}u_x+U_{ux})$,
resulting in the determining equation
\[
U_{uu}u_x+U_{ux}=0,
\]
which after splitting with respect to~$u_x$ leads to the representation
\[
U=U^1(t,y)u+U^0(t,x,y).
\]
Here, $U^1$ and $U^0$ are arbitrary smooth functions of $(t,y)$ and $(t,x,y)$, respectively.

Let us show that there are no more third- or fourth-order jets of $\tilde u$ appearing in the invariance condition.
Since we have already accounted for $\tilde u_{\tilde x\tilde x\tilde x\tilde x}$ and $\tilde u_{\tilde x\tilde x\tilde x}$,
the other third- or fourth-order parametric derivatives of $\tilde u$ would definitely contain $\tilde t$ or $\tilde y$ among indices.
Such derivatives originate only from $u_{xxxx}$, therefore their coefficients are proportional to $T_x$ or $Y_x$,
which are both zero.
This is why we continue to split the invariance condition with respect to jets up to the second order.

Since ${\rm D}_xT=0$, the jet $\tilde u_{\tilde t\tilde t}$
does not appear in the expressions for $u_{tx}$ and $u_{xx}$ in terms of derivatives of $\tilde u$
with respect to $\tilde t$, $\tilde x$ and $\tilde y$.
Therefore, $\tilde u_{\tilde t\tilde t}$ comes only from $u_{yy}$.
This gives us the determining equation $T_y=0$.
Differentiating the equation~\eqref{eq:RelationTX} with respect to $y$, we obtain $X^1_y=0$.

The jets $\tilde u_{\tilde x\tilde y}$ and $\tilde u_{\tilde y\tilde y}$ originate from the expressions of $u_{tx}$ and $u_{yy}$.
Respectively collecting coefficients at these derivatives, we obtain the equations
\begin{gather}\label{eq:ConstraintsOnX1andY1}
X^1Y_t=\frac32\sigma X^0_yY_y\quad\mbox{and}\quad
X^1T_t=Y_y^2.
\end{gather}
Differentiating the second equation of the system~\eqref{eq:ConstraintsOnX1andY1} with respect to $y$,
we obtain $Y_{yy}=0$, which integrates to $Y=Y^1(t)y+Y^0(t)$,
where $Y^1$ and $Y^0$ are arbitrary smooth functions of $t$.
The equation~\eqref{eq:RelationTX} implies $(Y^1)^2=T_t^{4/3}$, thus $Y^1=\varepsilon T_t^{2/3}$, where $\varepsilon=\pm1$.
The first equation of~\eqref{eq:ConstraintsOnX1andY1} gives us
\begin{gather}\label{eq:ConstraintX0}
X^0_y=\frac{2X^1Y_t}{3\sigma Y_y}=\frac49\sigma T_{tt}T_t^{-2/3}y+\frac23\sigma\varepsilon T_t^{-1/3}Y^0_t,
\end{gather}
where we used $\sigma^{-1}=\sigma$ since $\sigma=\pm1$.

Separately collecting coefficients of jet monomials $\tilde u\tilde u_{\tilde x\tilde x}$ and $\tilde u_{\tilde x\tilde x}$,
we derive the determining equations
\begin{gather}\label{eq:ConstratintU}
U^1=\frac{X^1}{T_t}\quad\mbox{and}\quad
\frac{U^0}{U^1}=\frac{2(X^1_tx+X^0_t)}{3X^1}-\frac{\sigma(X^0_y)^2}{2(X^1)^2}.
\end{gather}
After this step, the invariance condition becomes an identity,
hence, there are no further constraints on the components of $\Phi$.
Combining the representations for the components 
$T=T(t)$, $X=X^1(t)x+X^0(t,y)$, $Y=Y^1(t)y+Y^0(t)$ and $U=U^1(t)y+U^0(t,y)$,
with the constraints \eqref{eq:RelationTX}--\eqref{eq:ConstratintU} (after integrating some of them),
we obtain the transformations of the form~\eqref{eq:PointSymGroup}.
This finishes the proof of Theorem~\ref{thm:PointSymGroupKP}.

\subsection{Algebraic method}\label{sec:PSGAlgMethod}

The underlying idea of the algebraic method was first suggested by Hydon \cite{hydo1998a,hydo2000A,hydo2000b}:
the pushforward $\Phi_*$ of any element $\Phi$ of the point-symmetry (pseudo)group $G$ of a system of partial differential equations
is an automorphism of the Lie invariance algebra~$\mathfrak g$ of this system, $\Phi_*\mathfrak g=\mathfrak g$.
This insight led to the development of two versions of the algebraic method: based on the {\it automorphisms} and on the {\it megaideals}.

The automorphism-based version of the algebraic method is based on the explicit description
of the automorphism group ${\rm Aut}(\mathfrak g)$ of the Lie invariance algebra $\mathfrak g$.
However, obtaining such a description is rarely possible in practice for complex systems.
In particular, the algebras may be infinite-dimensional, like for the Kadomtsev--Petviashvili equation~\eqref{eq:KP}

The megaideal-based version~\cite{bihlo2011b} offers more flexibility
since its core principle consists in replacing the condition $\Phi_*\mathfrak g=\mathfrak g$
by $\Phi_*\mathfrak m\subseteq\mathfrak m$ for any megaideal $\mathfrak m\subseteq\mathfrak g$.
This modification allows even for finding symmetry (pseudo)groups when the Lie-symmetry algebra is infinite-dimensional.

To apply the megaideal-based version of the algebraic method to a system $\mathcal L$
of partial differential equations, one should first find sufficiently many megaideals $\mathfrak m$
of the maximal Lie invariance algebra $\mathfrak g$ of $\mathcal L$.
Then using the condition $\Phi_*\mathfrak m\subset\mathfrak m$,
determine constraints on the components of the point transformation $\Phi$.
To verify that no more constraints can be found within the framework of the algebraic method,
one should verify that $\Phi_*\mathfrak g=\mathfrak g$.
Finally, apply the direct method to find the remaining constraints
or to confirm that the symmetry group has been fully determined.
For more detailed description and applications of both automorphism- and megaideal-based versions
of the algebraic method see, e.g., \cite{bihl2011a,bihlo2011b,boyk2024a,kova2026c,kova2025a,vane2021a}
and references therein.

We are now ready to apply the megaideal-based version of the algebraic method
for proving Theorem~\ref{thm:PointSymGroupKP}.
Consider the most general form of the point transformation
acting on the space of the independent and dependent variables $\mathbb R_{t,x,y}^3\times\mathbb R_u$,
\begin{gather*}
\Phi\colon\quad(\tilde t,\tilde x,\tilde y,\tilde u)=(T,X,Y,U)(t,x,y,u)
\end{gather*}
with the nonzero Jacobian, $|\p(T,X,Y,U)/\p(t,x,y,u)|\ne0$.
In Section~\ref{sec:LieInvAlgebra}, we constructed the following chain of megaideals of the Lie algebra $\mathfrak g$:
\[
\mathfrak m_1=\mathfrak g=\langle\mathcal T(\tau),\mathcal X(\chi),\mathcal Y(\rho)\rangle,\quad
\mathfrak m_2=\langle \mathcal X(\chi),\mathcal Y(\rho)\rangle,\quad
\mathfrak m_3=\langle \mathcal X(\chi)\rangle,
\]
where $\tau$, $\chi$ and $\rho$ run through the space of smooth functions depending on $t$.
We use the conditions $\Phi_*\mathfrak m_i\subseteq\mathfrak m_i$
to derive principal constraints on the components of the point transformation~$\Phi$.

First, we observe that the $t$-components of vector fields from $\mathfrak m_3$ and $\mathfrak m_2$ are zero,
while the other components remain nonzero,
which immediately implies
\begin{gather}\label{eq:AlgMetConstOnT}
T_x=T_y=T_u=0,
\end{gather}
that is, $\tilde t=T(t)$.

Now consider the vector field $\mathcal X(\chi)\in\mathfrak m_3$,
where $\chi$ is an arbitrary smooth function.
The condition $\Phi_*(\mathcal X(\chi))\in\mathfrak m_3$
implies
\begin{gather*}
\Phi_*\big(\mathcal X(\chi)\big)=\mathcal X(\tilde\chi^1),
\end{gather*}
for some smooth function $\tilde\chi^1(t)$.
Composing both sides of this equation with $\Phi$ leads to the system of constraints on $X$, $Y$ and $U$:
\begin{subequations}
\begin{gather}
\label{eq:AlgMetDetForY}
\chi Y_x+\tfrac23\chi_tY_u=0,
\\ \label{eq:AlgMetCompatibilityForM3}
\chi X_x+\tfrac23\chi_tX_u=\tilde\chi^1(\tilde t),\quad
\chi U_x+\tfrac23\chi_tU_u=\tfrac23\tilde\chi^1_{\tilde t}(\tilde t).
\end{gather}
\end{subequations}
Since $\chi$ is an arbitrary smooth function,
we can split the equation~\eqref{eq:AlgMetDetForY} with respect to~$\chi$ and~$\chi_t$
(which can be done by setting, e.g., $\chi=1$ and $\chi=t$),
which yields
\begin{gather}\label{eq:AlgMetConstOnY}
Y_x=Y_u=0.
\end{gather}
After splitting the compatibility condition between the equations~\eqref{eq:AlgMetCompatibilityForM3}
with respect to~$\chi$ and~$\chi_t$,
we obtain
\begin{gather}\label{eq:AlgMetConstOnX}
X_u=0,\quad
\tfrac23X_{xt}=T_tU_x,\quad
T_tU_u=X_x.
\end{gather}
Furthermore, since $\tilde t=T(t)$, the following constraints hold:
\[
X_{xx}=X_{xy}=0,\quad
U_{uu}=U_{ux}=U_{uy}=0,\quad
U_{xx}=U_{xy}=0.
\]

The pushforward of vector field $\mathcal Y(\rho)\in\mathfrak m_2$ takes the form
\begin{gather*}
\Phi_*(\mathcal Y(\rho))=\mathcal Y(\tilde\rho^2)+\mathcal X(\tilde\chi^2),
\end{gather*}
where $\tilde\rho^2$ and $\tilde\chi^2$ are smooth function of $t$.
Composing both sides of this equality with $\Phi$ leads to the following system of constraints:
\begin{subequations}
\begin{gather}\label{eq:AlgMetYyyIsZero}
\rho Y_y=\tilde\rho^2(\tilde t)
\\\label{eq:AlgMetCompatibilityXY}
y\rho_tX_x+\tfrac32\sigma\rho X_y
=\tilde\chi^2(\tilde t)+\tilde y\tilde\rho^2_{\tilde t}(\tilde t)
\\\label{eq:AlgMetCompatibilityXY1}
y\rho_tU_x+\tfrac32\sigma\rho U_y+\tfrac23y\rho_{tt}U_u
=\tfrac23(\tilde\chi^2_{\tilde t}(\tilde t)+\tilde y\tilde\rho^2_{\tilde t\tilde t}(\tilde t))
\end{gather}
\end{subequations}
The equation~\eqref{eq:AlgMetYyyIsZero} gives us $Y_{yy}=0$.
Finding the compatibility condition between~\eqref{eq:AlgMetCompatibilityXY} and~\eqref{eq:AlgMetCompatibilityXY1}
gives us
\begin{gather}\label{eq:AlgMetConstOnX1}
\sigma X_{yt}-\frac23\frac{Y_t}{T_t}Y_{ty}=\frac32\sigma T_tU_y
\quad\mbox{and}\quad
\sigma X_y=\frac23\frac{Y_t}{T_t}Y_y.
\end{gather}
Expressing $\tilde\rho^2_{\tilde t}$ from~\eqref{eq:AlgMetYyyIsZero} using chain rule,
substituting it into~\eqref{eq:AlgMetCompatibilityXY} and differentiating the result with respect to $y$
results in the following constraint:
\begin{gather}\label{eq:AlgMetConstOnY1}
X_x=\frac{Y_y^2}{T_t}.
\end{gather}

Finally, pushing forward the vector field $\mathcal T(\tau)\in\mathfrak m_1$ gives
\begin{gather*}
\Phi_*\big(\mathcal T(\tau)\big)=
\mathcal T\big(\tilde\tau^3\big)+\mathcal Y\big(\tilde\rho^3\big)+\mathcal X\big(\tilde\chi^3\big)
\end{gather*}
for some smooth functions $\tilde\tau^3(t)$, $\tilde\rho^3(t)$ and $\tilde\chi^3(t)$.
Composing both sides of this equality with $\Phi$ yields the following system:
\begin{subequations}
\begin{gather}\label{eq:AlgMetConditionOnT}
\tau T_t=\tilde\tau^3(\tilde t),
\\\label{eq:AlgMetCompatinbilityYT}
\tau X_t+\tfrac13\Big(\tfrac23\sigma y^2\tau_{tt}+x\tau_t\Big)X_x+\tfrac23y\tau_tX_y
=\tfrac13t\Big(\tfrac23\sigma\tilde y^2\tilde\tau^3_{\tilde t\tilde t}(\tilde t)+\tilde x\tilde\tau^3_{\tilde t}(\tilde t)\Big)+\tilde y\tilde\rho^3_{\tilde t}(\tilde t)+\tilde\chi^3(\tilde t),
\\\label{eq:AlgMetYF}
\tau Y_t+\tfrac23y\tau_tY_y
=\tfrac23\tilde y\tilde\tau^3_{\tilde t}(\tilde t)+\tfrac32\sigma\tilde\rho^3(\tilde t),
\\\label{eq:AlgMetCompatinbilityYT1}
\begin{split}
&\tau U_t+\tfrac13\Big(\tfrac23\sigma y^2\tau_{tt}+x\tau_t\Big)U_x+\tfrac23y\tau_tU_y+\tfrac23\Big(\tfrac29\sigma y^2\tau_{ttt}-u\tau_t+\tfrac13x\tau_{tt}\Big)U_u
\\
&\qquad
=\tfrac23\Big(
\tfrac29\sigma\tilde y^2\tilde \tau^3_{\tilde t\tilde t\tilde t}(\tilde t)+\tfrac13\tilde x\tilde\tau^3_{\tilde t\tilde t}(\tilde t)
+\tilde y\tilde\rho^3_{\tilde t\tilde t}(\tilde t)+\tilde\chi^3_{\tilde t}(\tilde t)
\Big)-\tfrac23\tilde u\tilde\tau^3_{\tilde t}(\tilde t).
\end{split}
\end{gather}
\end{subequations}
Expressing the derivative $\tilde\tau^3_{\tilde t}$ from~\eqref{eq:AlgMetConditionOnT} using chain rule
and substituting it into~\eqref{eq:AlgMetYF}
and differentiating the result with respect to $y$ leads us to the constraint
\begin{gather}\label{eq:AlgMetConstOnY2}
Y_{ty}=\frac23\frac {Y_y}{T_t}T_{tt}.
\end{gather}
Integrating the system of constraints  constituted by
\eqref{eq:AlgMetConstOnT}, \eqref{eq:AlgMetConstOnY}, \eqref{eq:AlgMetConstOnX}, \eqref{eq:AlgMetConstOnX1},
\eqref{eq:AlgMetConstOnY1} and \eqref{eq:AlgMetConstOnY2}
we obtain the following form for a point transformations $\Phi$:
\begin{gather*}
\tilde t=T(t),\quad
\tilde x=c^2T_t^{1/3}x+\sigma c^2\frac{2T_{tt}}{9T_t^{2/3}}y^2
+\sigma c\frac{2Y^0_t}{3T_t^{1/3}}y+X^0(t),\quad
\tilde y=cT_t^{2/3}y+Y^0(t)
\\
\tilde u=\frac{c^2}{T_t^{2/3}}u+\frac{2T_{tt}}{9T_{t}^{5/3}}c^2x
+\frac{4\sigma c}{9T_t^{1/3}}\left(\frac{T_{tt}}{3T_t^{4/3}}cy^2+\frac{Y^0_t}{T_t}y\right)_t+U^0(t).
\end{gather*}
The compatibility condition for equations~\eqref{eq:AlgMetCompatinbilityYT} and~\eqref{eq:AlgMetCompatinbilityYT1}
allows us to specify the function $U^0$,
\[
U^0=\frac2{9T_t^2}\big(3 T_tX^0_t-\sigma(Y^0_t)^2\big).
\]
Note that for the correct consideration of the compatibility condition for~\eqref{eq:AlgMetCompatinbilityYT} and~\eqref{eq:AlgMetCompatinbilityYT1},
one should express the involved functions and their derivatives with tilde $\tilde\tau_{\tilde t\tilde t}$, $\tilde\tau_{\tilde t}$, $\tilde\tau$,
$\tilde\rho_{\tilde t}$, $\tilde\rho$ in terms of their counterparts without tildes using equations~\eqref{eq:AlgMetConditionOnT} and~\eqref{eq:AlgMetYF}.

Under the obtained restrictions on $\Phi$, we have that $\Phi_*\mathfrak g\subseteq\mathfrak g$,
thus no further constraints can be found within the framework of the algebraic method.

We now proceed with applying the direct method to the specific form of the point transformation
found previously by the algebraic method.
Pushing forward the equation~\eqref{eq:KP} by this transformation,
we obtain the last condition $c^2=1$, implying $c=\pm 1$.
This completes the proof of Theorem~\ref{thm:PointSymGroupKP}.

\noprint{
\begin{remark}
In the course of finding the pseudogroup $G$ using the algebraic method,
instead of applying the condition $\Phi_*Q\in\mathfrak m$ for vector fields $\mathcal X(\chi)$, $\mathcal Y(\rho)$ and $\mathcal T(\tau)$ 
with unspecified parameter-functions $\tau$, $\rho$ and $\chi$,
we could have restricted the consideration only to the following six linearly independent vector fields:
\[
\mathcal T(1),\quad
\mathcal T(t),\quad
\mathcal X(1),\quad
\mathcal X(t),\quad
\mathcal Y(1),\quad
\mathcal Y(t).
\]
These vector fields span a subalgebra $\mathfrak s\subset\mathfrak g$.
\todo Is this a subalgebra defining diffeomorphisms that stabilize $\mathfrak g$?
\end{remark}
}

\section{Contact-symmetry pseudogroup}\label{sec:ContactSyms}

Since the contact-symmetry algebra $\mathfrak g_c$ of the equation~\eqref{eq:KP}
is merely the first prolongation of the Lie symmetry algebra $\mathfrak g$,
$\mathfrak g_c=\mathfrak g_{(1)}$,
the megaideal structure of $\mathfrak g_c$ is the same as for $\mathfrak g$.
In particular, we straightforwardly see that $\mathfrak g_c$ has the following chain of megaideals:
\[
\mathfrak m_{1(1)}\varsupsetneq\mathfrak m_{2(1)}\varsupsetneq\mathfrak m_{3(1)},
\]
where $\mathfrak m_{i(1)}$ is the first prolongation of the megaideal $\mathfrak m_i\subset\mathfrak g$,
$i\in\{1,2,3\}$.

\begin{corollary}
The contact-symmetry pseudogroup $G_c$ of the Kadomtsev--Petviashvili equation~\eqref{eq:KP}
is the first prolongation $G_{(1)}$ of the pseudogroup $G$.
\end{corollary}

\begin{proof}
We use the same modification of the algebraic method as presented in Section~\ref{sec:PSGAlgMethod},
but replace the Lie invariance algebra $\mathfrak g$ with the contact algebra $\mathfrak g_c$
and the point transformation $\Phi$ with the contact transformation
\begin{gather*}
\Psi\colon\quad
(\tilde t,\tilde x,\tilde y,\tilde u, \tilde u_{\tilde t},\tilde u_{\tilde x},\tilde u_{\tilde y})
=(Z^t,Z^x,Z^y,U,U^t,U^x,U^y).
\end{gather*}
The right-hand side is a tuple constituted by smooth functions depending on $(t,x,y,u,u_t,u_x,u_y)$
with a nonvanishing Jacobian.
These functions also satisfy the contact condition
\begin{gather}\label{eq:ContactConditions}
(Z^\mu_\nu+Z^\mu_uu_\nu)U^\mu=U_\nu+U_uu_\nu,\quad
Z^\mu_{u_\nu}U^\mu=U_{u_\nu},
\end{gather}
where the indices $\mu,\nu$ run through the set $\{t,x,y\}$
and the summation with repeating indices is assumed.
A necessary condition for $\Psi$ to define a contact symmetry of~\eqref{eq:KP}
is that the push-forward $\Psi_*$ of contact vector fields by $\Psi$
satisfies $\Psi_*\mathfrak m_{i(1)}\subseteq\mathfrak m_{i(1)}$ for all $i\in\{1,2,3\}$.
Given that the computations are analogous to those presented in Section~\ref{sec:PSGAlgMethod},
we omit the intermediate steps and present only the essential constraints on the components of $\Psi$ which are derived from
the above necessary conditions.

The condition $\Psi_*(\mathcal X(\chi))\in\mathfrak m_{3(1)}$ leads to $Z^\mu_{u_t}=0$ and $Z^\mu_u=0$ for all $\mu\in\{t,x,y\}$.
Next, the condition $\Psi_*(\mathcal Y(\rho))\in\mathfrak m_{2(1)}$ results in
$Z^\mu_{u_y}=0$, $\mu\in\{t,x,y\}$, and $Z^t_{u_x}=Z^y_{u_x}=0$.
Finally, $\Psi_*(\mathcal T(\tau))\in\mathfrak m_{1(1)}$ gives $Z^x_{u_x}=0$.
Consequently, the second equation in the contact condition~\eqref{eq:ContactConditions} leads to $U_{u_\nu}=0$.
This demonstrates that a transformation $\Psi$ is necessarily a prolongation of a point transformation 
in the fibred space $\mathbb R^3_{(t,x,y)}\times\mathbb R_u$.
\end{proof}

Corollary~\ref{eq:DiscreteSymmetriesOfKP} on the discrete elements of the point-symmetry pseudogroup $G$
extends {\it mutatis mutandis} to the discrete contact symmetries of~\eqref{eq:KP}.

\section{One- and two-dimensional subalgebras}\label{sec:Subalgebras}

To efficiently carry out Lie reductions of the Kadomtsev--Petviashvili equation~\eqref{eq:KP},
it suffices to classify one- and two-dimensional subalgebras of the maximal Lie invariance algebra~$\mathfrak g$,
see, for example, the discussion in~\cite[Section~2]{vinn2024a}.
In the seminal paper~\cite{davi1986a} on the symmetry analysis of the latter equation,
the one-, two- and three-dimensional subalgebras of~$\mathfrak g$
were classified up to the equivalence arising from the action by inner automorphisms of~$\mathfrak g$
on the algebra $\mathfrak g$.
While this is the classical approach to such problems, it typically requires solving complex differential and functional equations.

In this section, we independently verify the classification of one- and two-dimensional subalgebras
modulo the action of the point symmetry pseudogroup~$G$ via pushforwards on the algebra~$\mathfrak g$.
This approach is more convenient (in particular, in the infinite-dimensional setting):
rather than leading to complicated differential and functional equations,
it relies entirely on the explicit form of the pseudogroup~$G$.
Furthermore, we do not restrict our analysis to the identity component of~$G$;
instead, we exploit the complete pseudogroup~$G$, including its discrete elements.

Thus, the nonidentity adjoint actions of elementary transformations from $G$
on the vector fields~\eqref{eq:LieInvAlgebra} spanning the algebra $\mathfrak g$ are exhausted by
\begin{gather*}
\mathscr T(T)_*\mathcal T(\tau)=\mathcal T\Big(\hat T_t^{-1}\tau(\hat T)\Big),\ 
\mathscr T(T)_*\mathcal X(\chi)=\mathcal X\Big(\hat T_t^{-1/3}\chi(\hat T)\Big),\ 
\mathscr T(T)_*\mathcal Y(\rho)=\mathcal Y\Big(\hat T_t^{-2/3}\rho(\hat T)\Big),
\\[0.5ex]
\mathscr Y(Y^0)_*\mathcal T(\tau)=\mathcal T(\tau)+\tfrac{2\sigma}3\mathcal Y\Big(\tau Y^0_t-\tfrac23\tau_tY^0\Big),\quad
\mathscr Y(Y^0)_*\mathcal Y(\rho)=\mathcal Y(\rho)+\mathcal X\Big(\rho Y^0_t-\rho_tY^0\Big),
\\[0.5ex]
\mathscr X(X^0)_*\mathcal T(\tau)=\mathcal T(\tau)+\mathcal X\Big(\tau X^0_t-\tfrac13X^0\tau_t\Big),
\\[0.5ex]
\mathscr I_*\mathcal Y(\rho)=-\mathcal Y(\rho),
\end{gather*}
where $\hat T$ denotes the inverse function of $T$.

\begin{theorem}\label{thm:1DSubalgebrasKP}
A complete list of $G$-inequivalent one-dimensional subalgebras of $\mathfrak g$
is exhausted by the following subalgebras:
\begin{gather*}
\mathfrak s_{1.1}=\langle\mathcal T(1)\rangle,\qquad
\mathfrak s_{1.2}=\langle\mathcal Y(1)\rangle,\qquad
\mathfrak s_{1.3}=\langle\mathcal X(1)\rangle.
\end{gather*}
\end{theorem}

\begin{theorem}\label{thm:2DSubalgebrasKP}
Any two-dimensional subalgebra of the Lie algebra~$\mathfrak g$
is equivalent to a member of the following list:

1. abelian subalgebras:
\begin{gather*}
\mathfrak s_{2.1}  =\langle\mathcal T(1),\mathcal Y(1)\rangle,\quad
\mathfrak s_{2.2}  =\langle\mathcal T(1),\mathcal X(1)\rangle,\quad
\mathfrak s_{2.3}^h=\langle\mathcal Y(1),\mathcal X(h)\rangle,\quad
\mathfrak s_{2.4}^f=\langle\mathcal X(1),\mathcal X(f)\rangle,
\end{gather*}
where $h$ and $f$ are arbitrary nonzero smooth function of $t$ with $f_t\ne0$.

2. nonabelian subalgebras:
\begin{gather*}
\mathfrak s_{2.5}=\langle\mathcal T(t),\mathcal T(1)\rangle,\quad
\mathfrak s_{2.6}=\langle\mathcal T\big(t\big),\mathcal Y(1)\rangle,\quad
\mathfrak s_{2.7}=\langle\mathcal T(t),\mathcal X(1)\rangle.
\end{gather*}
Furthermore, in the above list, the only $G$-equivalent pairs of subalgebras are
\begin{gather*}
\mathfrak s_{2.3}^{h^1}\sim\mathfrak s_{2.3}^{h^2}
\quad\text{if}\quad h^2(t)=a\,h^1(t-b)
\quad\mbox{for some}\ a\in\mathbb R\setminus\{0\}, b\in\mathbb R\ \mbox{and}
\\
\mathfrak s_{2.4}^{h^1}\sim\mathfrak s_{2.4}^{h^2}
\quad\text{if}\quad
h^2\big(\tau(t)\big)
=\frac{d_{11}h^1(t)-d_{21}}{d_{22}-d_{12}h^1(t)}\quad\mbox{with}\quad
\tau(t):=\int\frac{\Delta^3{\rm d}t}{\big(d_{22}-d_{12}h^1(t)\big)^3},
\end{gather*}
where $d_{11}$, $d_{12}$, $d_{21}$ and $d_{22}$ are arbitrary real constants satisfying $\Delta:=d_{11}d_{22}-d_{12}d_{21}\ne0$.
\end{theorem}

\begin{remark}
The equivalence relations within the family of subalgebras~$\mathfrak s_{2.4}$ 
established in the previous theorem were omitted in~\cite{davi1986a}. 
Nevertheless, this omission does not affect the resulting Lie reductions or the construction of Lie invariant solutions.
\end{remark}

\section{Codimension-one reductions}\label{sec:Codim1Reductions}

We study reduced invariant submodels of the Kadomtsev--Petviashvili equation~\eqref{eq:KP}
following the algorithm presented in \cite[Section~2]{vinn2024a}.
The classification of codimension-one Lie reductions of~\eqref{eq:KP}
is based on the list of one-dimensional $G$-inequivalent subalgebras of the Lie invariance algebra~$\mathfrak g$.
To each one-dimensional subalgebra from the list in Theorem~\ref{thm:1DSubalgebrasKP}
we construct an appropriate ansatz and derive the corresponding reduced equation.
Throughout this section, numerical subscripts 1 and 2 on functions of $(z_1,z_2)$ denote partial derivatives
with respect to~$z_1$ and~$z_2$, respectively.

We also compute the Lie invariance algebras and point-symmetry pseudogroups of each reduced model,
leading to a comprehensive description of the induced (and, where appropriate, hidden)
infinitesimal and point symmetries of both the reduced and original systems.
Although it is suggested in \cite[Section~2]{vinn2024a} that two-step reductions should be carried out
using only hidden infinitesimal symmetries of~\eqref{eq:KP}, we perform all two-step reductions,
revisiting the classical symmetry analysis of the reduced models in detail.
We account for the overlap of such two-step reductions with the codimension-two Lie reductions of~\eqref{eq:KP} in Section~\ref{sec:Codim2Reductions}.

\subsection{Subalgebra family 1.1}\label{sec:Reduction11}

A suitable ansatz constructed using the first subalgebra $\mathfrak s_{1.1}=\langle\p_t\rangle$
is $u=\tfrac\sigma6v(z_1,z_2)$ with $z_1=y/\sqrt3$ and $z_2=x$,
and it reduces the original equation~\eqref{eq:KP} to the Boussinesq equation:
\begin{equation}\label{eq:Red11Boussinesq}
v_{11}=\tfrac12(v^2)_{22}+\sigma v_{2222}.
\end{equation}

\paragraph{Maximal Lie invariance algebra.}
The maximal Lie invariance algebra~$\mathfrak g_{1.1}$ of~\eqref{eq:Red11Boussinesq}
was computed in~\cite{clar1989a}.
It is the span of the following vector fields:
\begin{gather}\label{eq:BoussinesqLieSyms}
D^z:=z_1\p_{z_1}+\tfrac12z_2\p_{z_2}-v\p_v,\quad
\p_{z_1},\quad
\p_{z_2}.
\end{gather}
Up to the skew-symmetry of the Lie bracket, the nonzero commutation relations
between vector fields~\eqref{eq:BoussinesqLieSyms} are exhausted by
\noprint{
\[
[D^z,\p_{z_1}]=-\p_{z_1},\quad
[D^z,\p_{z_2}]=-\tfrac12\p_{z_2}.
\]}
\[
[\p_{z_1},D^z]=\p_{z_1},\quad
[\p_{z_2},D^z]=\tfrac12\p_{z_2}.
\]
Therefore, the Lie algebra $\mathfrak g_{1.1}$ is isomorphic to the
three-dimensional almost abelian Lie algebra with the semisimple element $\mathop{\rm diag}(-1,-\tfrac12)$,
$\mathfrak g_{1.1}\simeq A_{3.4}^{1/2}$.

The normalizer of the subalgebra $\mathfrak s_{1.1}$ in $\mathfrak g$ is the span
$\langle t\p_t+\tfrac13x\p_x+\tfrac23y\p_y-\tfrac23u\p_u,\p_t,\p_x,\p_y\rangle$,
hence the algebra $\mathfrak g_{1.1}$ is entirely induced from $\mathfrak g$.
In other words, there are no hidden (infinitesimal) symmetries of the Kadomtsev--Petviashvili equation
with respect to the reduction by the subalgebra $\mathfrak s_{1.1}$.

\paragraph{Point symmetry group}
\begin{theorem}\label{thm:PSGBoussinesq}
The point symmetry group $G_{1.1}$ of the Boussinesq equation~\eqref{eq:Red11Boussinesq}
is constituted by the transformations of the form:
\begin{gather}\label{eq:PSGBoussinesq}
\tilde z_1=\epsilon\alpha^2z_1+\lambda_1,\quad
\tilde z_2=\alpha z_2+\lambda_2,\quad
\tilde v  =\alpha^{-2}v,
\end{gather}
where $\alpha$, $\lambda_0$ and $\lambda_1$ are arbitrary real constants with $\alpha\ne0$
and $\epsilon=\pm 1$.
\end{theorem}

\begin{proof}
To compute the group $G_{1.1}$ we use Hydon's automorphism-based version of the algebraic method \cite{hydo1998a,hydo2000A,hydo2000b}.

The isomorphism $\mathfrak g_{1.1}\simeq A_{3.4}^{1/2}$ is explicitly given by the correspondence
between the elements of the bases
$(\p_{z_1},\p_{z_2},D^z)\mapsto(e_1,e_2,e_3)$ and is extended by linearity.
Therefore, in the basis $(\p_{z_1},\p_{z_2},D^z)$ of the Lie algebra $\mathfrak g_{1.1}$,
the automorphism group ${\rm Aut}(\mathfrak g_{1.1})$ consists of the matrices
\begin{gather*}
\begin{pmatrix}
\alpha_{11}&0&\alpha_{13}\\
0&\alpha_{22}&\alpha_{23}\\
0&0&1
\end{pmatrix},
\end{gather*}
where the parameters $\alpha_{11}$, $\alpha_{22}$, $\alpha_{13}$ and $\alpha_{23}$
are arbitrary real constants with $\alpha_{11}\alpha_{22}\ne0$, see, e.g.,~\cite{popo2003a}.

Consider an arbitrary point transformation $\Phi$ acting on the space of independent and dependent variables
$\mathbb R^2_{z_1,z_2}\times\mathbb R_v$,
\[
\Phi\colon\quad
(\tilde z_1,\tilde z_2,\tilde v)=(Z^1,Z^2,V)(z_1,z_2,v),
\]
with a nonvanishing Jacobian, $|\p(Z^1,Z^2,V)/\p(z_1,z_2,v)|\ne0$.
If $\Phi$ determines a point symmetry of the Boussinesq equation~\eqref{eq:Red11Boussinesq},
then the pushforward $\Phi_*$ is an automorphism of $\mathfrak g_{1.1}$.
The explicit matrix representation of ${\rm Aut}(\mathfrak g_{1.1})$ leads to the necessary conditions
\begin{gather*}
\Phi_*(\p_1)=\alpha_{11}\p_{z_1},\quad
\Phi_*(\p_2)=\alpha_{22}\p_{z_2},\quad
\Phi_*(D^z) =\alpha_{13}\p_{z_1}+\alpha_{23}\p_{z_2}+D^z.
\end{gather*}
This yields a system of determining equations on the components of $\Phi$:
\begin{gather*}
Z^1_{z_1}=\alpha_{11},\quad
Z^2_{z_1}=V_{z_1}=0,\quad
Z^2_{z_2}=\alpha_{22},\quad
Z^1_{z_2}=V_{z_2}=0,
\\
z_1Z^1_{z_1}-vZ^1_v=Z^1+\alpha_{13},\quad
\tfrac12z_2Z^2_{z_2}-vZ^2_v=\tfrac12Z^2+\alpha_{23},\quad
vV_v=V,
\end{gather*}
whose general solution is 
\begin{gather}\label{eq:GenSolAlgebraicMForBoussinesq}
Z^1=\alpha_{11}z_1+\beta_1v^{-1}-\alpha_{13},\quad
Z^2=\alpha_{22}z_2+\beta_2v^{-1/2}-2\alpha_{23},\quad
V=\beta_3v,
\end{gather}
where $\beta_1$, $\beta_2$ and $\beta_3$ are arbitrary real constants
with $\beta_3\ne0$.
Next, we apply the direct method to the equation~\eqref{eq:Red11Boussinesq}
using the expression~\eqref{eq:GenSolAlgebraicMForBoussinesq} as the general form for components of $\Phi$.
Analogously to the procedure in Section~\ref{sec:PSGDirectMethod}, we express $v$ and its derivatives with respect to $z_1$ and $z_2$
via the new dependent variable $\tilde v$ and its derivatives with respect to
the new independent variables $\tilde z_1$ and $\tilde z_2$ using the chain rule,
substitute the obtained expressions into~\eqref{eq:GenSolAlgebraicMForBoussinesq}
and require the resulting equation to be satisfied for all solutions of~\eqref{eq:GenSolAlgebraicMForBoussinesq}. 
First, we note that collecting coefficients at $\tilde v_{\tilde z_1\tilde z_1\tilde z_1\tilde z_1}$
and $\tilde v_{\tilde z_2\tilde z_2\tilde z_2\tilde z_2}$
respectively results in the equations
\[
({\rm D}_{z_2}Z^1)^4=0,\quad
({\rm D}_{z_1}Z^1)^2-({\rm D}_{z_2}Z^2)^4=0.
\]
This yields the constraints $\beta_1=\beta_2=0$ and $|\alpha_{11}|=\alpha_{22}^2$.
Such simplification allows to write the transformed jet monomials that appear in~\eqref{eq:Red11Boussinesq} as follows:
\[
\tilde v_{\tilde z_1\tilde z_1}=\beta_3\alpha_{22}^{-4}v_{11},\quad
(\tilde v)^2_{\tilde z_2\tilde z_2}=\beta_3^2\alpha_{22}^{-2}(v^2)_{22},\quad
\tilde v_{\tilde z_2\tilde z_2\tilde z_2\tilde z_2}=\beta_3\alpha_{22}^{-4}v_{2222}.
\]
Substituting these expressions into~\eqref{eq:Red11Boussinesq}
leads to the constraint $\beta_3=\alpha_{22}^{-2}$.
Relabeling the parameters according to the rule $(\alpha_{22},\alpha_{13},\alpha_{23})\mapsto(\alpha,-\lambda_1,-\tfrac12\lambda_2)$
results in the statement of the theorem.
\end{proof}

The group $G_{1.1}$ is the semidirect product of its subgroup consisting of transformations
of the form~\eqref{eq:PSGBoussinesq} with $\lambda_1=\lambda_2=0$
by its normal subgroup with $\alpha=\epsilon=1$.
Therefore, $G_{1.1}$ is isomorphic to the Lie group $(\mathbb Z_2\times\mathbb R^\times)\ltimes\mathbb R^2$,
where the action of $\mathbb Z_2\times\mathbb R^\times$ on the normal subgroup $\mathbb R^2$
is given by $(\epsilon,\alpha)\cdot(\lambda_1,\lambda_2)=(\epsilon\alpha^2\lambda_1,\alpha\lambda_2)$.
The identity component of $G_{1.1}$ is constituted by the transformations~\eqref{eq:PSGBoussinesq}
with $\epsilon=1$ and $\alpha>0$,
hence a complete list of discrete symmetries of the Boussinesq equation~\eqref{eq:Red11Boussinesq}
that are independent up to composing them with each other and with elements of the identity component of  $G_{1.1}$
are exhausted by two commuting involutions of alternating signs of $z_1$ and $z_2$,
$(z_1,z_2,v)\mapsto(-z_1,z_2,v)$ and $(z_1,z_2,v)\mapsto(z_1,-z_2,v)$.
They generate the group isomorphic to the Klein four-group $\mathbb Z_2\times\mathbb Z_2$.

For a point symmetry $\Phi$ of the Kadomtsev--Petviashvili equation~\eqref{eq:KP} 
to induce a point symmetry of the Boussinesq equation~\eqref{eq:Red11Boussinesq},
it must preserve the foliation of $\mathbb R^4_{t,x,y,u}$ cut out by the equations $x=c_1$, $y=c_2$ and $u=c_3$,
where $c_1$, $c_2$ and $c_3$ are arbitrary real constants.
Therefore, the components of $\Phi$  given in~\eqref{eq:PointSymGroup} satisfy the differential equations 
\[
T_{tt}=0,\quad
X^0_t=0\quad\mbox{and}\quad
Y^0_t=0.
\]
These constraints yield the point transformations
\[
\tilde t=\tau_1 t+\tau_0,\quad
\tilde x=\tau_1^{1/3}x+x_0,\quad
\tilde y=\varepsilon\tau_1^{2/3}y+y_0,\quad
\tilde u=\tau_1^{-2/3}u,
\]
where $\tau_1$, $\tau_0$, $x_0$ and $y_0$ are arbitrary real constants with $\tau_1\ne0$,
and $\varepsilon=\pm1$.
This implies that all point symmetries of the Boussinesq equation~\eqref{eq:Red11Boussinesq} are entirely induced
from the point symmetries of the  Kadomtsev--Petviashvili equation~\eqref{eq:KP}.

\paragraph{Lie reductions.}

Some Lie reductions of the Boussinesq equation~\eqref{eq:Red11Boussinesq} were carried out in~\cite{clar1989a}
using the vector fields of the form $\alpha D^z+\beta\p_{z_1}+\gamma\p_{z_2}$ with $\alpha=0$ and $\alpha\ne0$,
where $\alpha$, $\beta$ and $\gamma$ are real constants.
This approach, based on the general form of a Lie symmetry vector field, typically leads to complex multi-parameter differential equations;
furthermore, it yields equivalent solutions that can be obtained by the action of the point symmetry group $G_{1.1}$ on a simple seed invariant solution.
These complications can be essentially avoided by using non-equivalent (simplest) one- and two-dimensional subalgebras
of the maximal Lie invariance algebra~$\mathfrak g_{1.1}$ under the action of $G_{1.1}$ by pushforwards.
See a more detailed discussion on the effectiveness of the latter approach in~\cite[Section~2]{vinn2024a}

\begin{proposition}
A complete list of $G_{1.1}$-inequivalent proper subalgebras of the maximal Lie invariance algebra~$\mathfrak g_{1.1}$
of the Boussinesq equation~\eqref{eq:Red11Boussinesq} is exhausted by the following subalgebras:
\begin{gather*}
{\rm 1D}\colon \
\langle\p_{z_1}\rangle,\quad
\langle\p_{z_2}\rangle,\quad
\langle D^z\rangle,\quad
\langle\p_{z_1}+\p_{z_2}\rangle,
\\
{\rm 2D}\colon \
\langle\p_{z_1},\p_{z_2}\rangle,\quad
\langle D^z,\p_{z_1}\rangle,\quad
\langle D^z,\p_{z_2}\rangle.
\end{gather*}
\end{proposition}

\begin{proof}
Under the isomorphism $\mathfrak g_{1.2}\simeq A_{3.4}^{1/2}$, we utilize the classification of proper subalgebras of $A_{3.4}^{1/2}$
with respect to the action of its group of inner automorphisms 
(which is equivalent to the action of the identity component of the group $G_{1.1}$ on $\mathfrak g_{1.1}$)
presented in~\cite{pate1977a}.
This classification is as follows: 
\begin{gather*}
\langle\p_{z_1}\rangle,\quad
\langle\p_{z_2}\rangle,\quad
\langle D^z\rangle,\quad
\langle\p_{z_1}+\varepsilon\p_{z_2}\rangle,
\quad
\langle\p_{z_1},\p_{z_2}\rangle,\quad
\langle D^z,\p_{z_1}\rangle,\quad
\langle D^z,\p_{z_2}\rangle,
\end{gather*}
where $\varepsilon=\pm1$.
The action by the discrete point symmetry $(z_1,z_2,v)\mapsto(-z_1,z_2,v)$
allows us to gauge the parameter $\varepsilon$ in the subalgebra $\langle\p_{z_1}+\varepsilon\p_{z_2}\rangle$ to $1$.
This is the only non-identity action of the discrete elements of $G_{1.1}$ on the list from~\cite{pate1977a}.
\end{proof}

We arrange the codimension-one Lie reductions of the Boussinesq equation with respect to
one-dimensional subalgebras listed in the previous lemma in Table~\ref{tab:LieRedBoussinesq1}.
For each subalgebra, we write a suitable ansatz for $v$ and the resulting reduced 
ordinary differential equation.

\begin{table}[!ht]\footnotesize
\begin{center}
\caption{$G_{1.1}$-inequivalent codimension-one Lie reductions of the Boussinesq equation}
\label{tab:LieRedBoussinesq1}
${ }$\\[-1ex]
\renewcommand{\arraystretch}{1.6}
\begin{tabular}{|l|l|c|c|l|l|}
\hline 
no. & \!Subalgebra\!                                 & $v$              & $\omega$    & \hfil Reduced equation \\
\hline
1. & $\langle\p_{z_1}\rangle$                       &$\varphi$         & $z_2$       & $\sigma\varphi_{\omega\omega\omega\omega}+\tfrac12(\varphi^2)_{\omega\omega}=0$\\
2. & $\langle\p_{z_2}\rangle$                       &$\varphi$         & $z_1$       & $\varphi_{\omega\omega}=0$ \\
3. & $\langle D^z\rangle$                           &$z_1^{-1}\big(\varphi+\frac{z_2^2}{4z_1}\big)$ & $z_2|z_1|^{-\tfrac12}$   &  $\sigma\varphi_{\omega\omega\omega\omega}+\tfrac12\epsilon(\varphi^2)_{\omega\omega}=\tfrac34\big((\omega\varphi)_\omega+\varphi\big)+\tfrac98\epsilon\omega^2$ \\
4. & $\langle\p_{z_1}+\p_{z_2}\rangle$              &$\varphi+1$         & $z_2-z_1$   & $\sigma\varphi_{\omega\omega\omega\omega}+\tfrac12(\varphi^2)_{\omega\omega}=0$ \\
\hline
\end{tabular}\\[2ex]
Here $\epsilon':=\sgn z_1$, $\varphi=\varphi(\omega)$ is the new unknown function of the invariant independent variable~$\omega$,
$\epsilon:=\sgn(z_1)$.
\end{center}
\end{table}

It is stated in~\cite{clar1989a} that reductions 1, 3 and 4 from Table~\ref{tab:LieRedBoussinesq1}
can be integrated in terms of Weierstrass functions, solutions of Painlev\'e I or IV.
Nevertheless, the authors did not specify explicit point transformations that relate these reductions
to the Weierstrass or Painlev\'e equations.
While finding such a transformation for reductions 1 and 4 is relatively straightforward,
doing so for case 3 is very difficult.
Moreover, in this latter case, such a point transformation is, in general, {\it complex}.

\medskip\par\noindent
{\bf Reductions 1 and 4.} Twice integrating the reduced equations 1 and 4
yields
\begin{gather}\label{eq:ReductionBoussinesq}
\sigma\varphi_{\omega\omega}+\tfrac12\varphi^2=A\omega+B,
\end{gather}
where $A$ and $B$ are arbitrary real constants.
When $A=0$, the equation~\eqref{eq:ReductionBoussinesq} reduces to the Weierstrass equation,
and its general solution is given by
\[
\varphi=-12\sigma\wp(\omega+c_1,\tfrac16B,c_2),
\]
where $\wp$ denotes the Weierstrass elliptic function and $c_1$ and $c_2$ are arbitrary constants.
The corresponding solutions of the Boussinesq equation are
\begin{gather*}
\solutionRedEq 
v=-12\sigma\wp(z_2+c_1,\tfrac16c_2,c_3),
\\
\solutionRedEq 
v=-12\sigma\wp(z_2-z_1+c_1,\tfrac16c_2,c_3)+1,
\end{gather*}
where $c_1$, $c_2$ and $c_3$ are arbitrary real constants.
This leads to the following solutions of the Kadomtsev--Petviashvili equation~\eqref{eq:KP}:
\begin{gather*}
\solution
u=-2\wp(x+c_1,\tfrac16c_2,c_3),
\\
\solution
u=-2\wp(x-y/\sqrt3+c_1,\tfrac16c_2,c_3)+\tfrac\sigma6.
\end{gather*}

If $A\ne0$ in~\eqref{eq:ReductionBoussinesq}, then the ansatz
\[
\varphi=-12\sigma\left(-\frac{A}{12}\right)^{2/5}\tilde\varphi(\tilde\omega)\quad\mbox{with}\quad
\tilde\omega=\left(-\frac{A}{12}\right)^{1/5}\left(\omega+\frac BA\right)
\]
yields the Painlev\'e I equation
\[
\tilde\varphi_{\tilde\omega\omega}=6\tilde\varphi^2+\tilde\omega.
\]
Hence, given a solution $\mathtt y_{\rm I}(\tilde\omega)$ of the Painlev\'e I equation,
the corresponding solutions of Boussinesq equation~\eqref{eq:Red11Boussinesq} are given by
\noprint{
\begin{gather*}
\solutionRedEq
v=-12\sigma\left(-\frac{A}{12}\right)^{2/5}\mathtt y_{\rm I}\Big(\left(-\frac{A}{12}\right)^{1/5}\left(z_2+\frac BA\right)\Big),
\\
\solutionRedEq
v=-12\sigma\left(-\frac{A}{12}\right)^{2/5}\mathtt y_{\rm I}\Big(\left(-\frac{A}{12}\right)^{1/5}\left(z_2-z_1+\frac BA\right)\Big)+1,
\end{gather*}
}
\begin{gather*}
\solutionRedEq
v=-12\sigma(-A/12)^{2/5}\mathtt y_{\rm I}\Big((-A/12)^{1/5}(z_2+B/A)\Big),
\\
\solutionRedEq
v=-12\sigma(-A/12)^{2/5}\mathtt y_{\rm I}\Big((-A/12)^{1/5}(z_2-z_1+B/A)\Big)+1,
\end{gather*}
leading to the following solutions of the Kadomtsev--Petviashvili equation:
\begin{gather*}
\solution
u=-2(-A/12)^{2/5}\mathtt y_{\rm I}\Big((-A/12)^{1/5}(x+B/A)\Big),
\\
\solution
u=-2(-A/12)^{2/5}\mathtt y_{\rm I}\Big(\left(-A/12\right)^{1/5}\left(x-y/\sqrt3+B/A\right)\Big)+\frac\sigma6,
\end{gather*}
where $A$ and $B$ are arbitrary real constants with $A\ne0$.

\medskip\par\noindent
{\bf Reduction 3.}
It is stated in~\cite{clar1989a} that the reduced equation arising from the subalgebra $\langle D^z\rangle$
\begin{gather}\label{eq:ReductionBoussinesq2}
\sigma\varphi_{\omega\omega\omega\omega}+\tfrac12\epsilon(\varphi^2)_{\omega\omega}
=\tfrac34\big((\omega\varphi)_\omega+\varphi\big)+\tfrac98\epsilon\omega^2
\end{gather}
can be further related to Painlev\'e IV.
This claim is somewhat imprecise.
More precisely, equation \eqref{eq:ReductionBoussinesq2} can be related
to the equation for the {\it Hamiltonian} associated with the Painlev\'e IV equation.

We proceed by following the derivations from \cite[Appendix A]{clar1989a}, 
specifying all necessary point substitutions that were not explicitly stated there.
Changing the independent variable $\omega\to\sqrt{\epsilon}\omega$,
substituting $\varphi=\psi_\omega+\tfrac34\omega^2$ into~\eqref{eq:ReductionBoussinesq2},
where $\psi=\psi(\omega)$, and integrating the result twice with respect to $\omega$
leads to
\begin{gather*}
\sigma\psi_{\omega\omega\omega}+\tfrac12(\psi_\omega)^2+\tfrac34\omega(\omega\psi_\omega-\psi)=c_1\omega+c_2,
\end{gather*}
where $c_1$ and $c_2$ are arbitrary real constants.
Multiplying the last equation by $\psi_{\omega\omega}$ and integrating the result once yields
\begin{gather*}
\tfrac12\sigma(\psi_{\omega\omega})^2+\tfrac16(\psi_\omega)^3+\tfrac38(\omega\psi_\omega-\psi)^2
=c_1(\omega\psi_\omega-\psi)+c_2\psi_\omega+c_3.
\end{gather*}
The substitution
\[
\psi=-18(-\sigma/3)^{3/4}\left(\tilde\psi(\tilde\omega)+\frac{2c_1}{27}(-\sigma/3)^{-3/4}\right)\quad\mbox{with}\quad
\tilde\omega=\frac{\omega}2(-\sigma/3)^{-1/4},
\]
which is complex, providing $\sigma=1$,
further reduces the latter equation to
\begin{gather}\label{eq:ChazyIV}
(\tilde\psi_{\tilde\omega\tilde\omega})^2+4(\tilde\psi_{\tilde\omega})^3-4(\tilde\omega\tilde\psi_{\tilde\omega}-\tilde\psi)^2
+\alpha\tilde\psi_{\tilde\omega}+\beta=0
\\\nonumber
\quad\mbox{with}\quad\alpha:=\frac{8c_2}{9\sigma},\quad
\beta:=-\frac8{81\sigma\sqrt{-3\sigma}}(2c_1^2+3c_3).
\end{gather}
The equation~\eqref{eq:ChazyIV} is referred in~\cite[Appendix A]{clar1989a} as Chazy IV equation,
which was the last step of their derivation. 
The substitution
\[
\tilde\psi=\hat\psi(\tilde\omega)+\lambda\tilde\omega
\]
where $\lambda$ is a root of the cubic equation $4\lambda^3+\alpha\lambda+\beta=0$,
further relates~\eqref{eq:ChazyIV} to the equation on Hamiltonian of Painlev\'e IV~\cite[Equation~(1.16)]{clar2019a}:
\begin{gather}\label{eq:HamiltPainleveIV}
(\hat\psi_{\tilde\omega\tilde\omega})^2-4(\tilde\omega\hat\psi_{\tilde\omega}-\hat\psi)^2
+4\hat\psi_{\tilde\omega}(\hat\psi_{\tilde\omega}+2\vartheta_0)(\hat\psi_{\tilde\omega}+2\vartheta_\infty)=0,
\\ \nonumber\quad
\mbox{where}\quad
\vartheta_0:=\tfrac14(3\lambda+\sqrt{-(3\lambda^2+\alpha)}),\quad
\vartheta_\infty:=\tfrac14(3\lambda-\sqrt{-(3\lambda^2+\alpha)}).
\end{gather}

More specifically, given two solutions $\mathtt p(z)$ and $\mathtt q(z)$ of Painlev\'e IV~\cite[Equation~(1.2)]{clar2014a}
with $A:=1-\vartheta_0+2\vartheta_\infty$, $B:=-2\vartheta_0^2$ and
$A:=-1+2\vartheta_0-\vartheta_\infty$, $B:=-2\vartheta_\infty^2$, respectively,
one can define the Hamiltonian $\mathtt h_{\rm IV}(z)$ of Painlev\'e IV by \cite[Equation~(3.3)]{clar2014a}:
\[
\mathtt h_{\rm IV}(z):=2\mathtt q(z)\mathtt p^2(z)-\big(\mathtt q^2(z)+z\mathtt q(z)+2\vartheta_0\big)\mathtt p(z)+\vartheta_\infty\mathtt q(z).
\]
The function $\mathtt h_{\rm IV}(\hat\omega)$ turns out to be a solution of~\eqref{eq:HamiltPainleveIV}. 
Pulling back $\mathtt h_{\rm IV}(\hat\omega)$, we obtain the following solution of the Boussinesq equation:
\begin{gather*}
\solutionRedEq
v=-9\frac{s^{1/2}}{z_1}\mathtt h_{\rm IV}'\left(\frac{z_2}{2s^{1/4}\sqrt{z_1}}\right)
-9\lambda\frac{s^{1/2}}{z_1}+\frac{z_2^2}{4z_1^2}\quad\mbox{with}\quad
s=-\sigma/3,
\end{gather*}
where $\lambda$ is an arbitrary constant
and $\mathtt h_{\rm IV}'(z)$ denotes the derivative of $\mathtt h_{\rm IV}(z)$ with respect to its argument.
This yields the following solution of the Kadomtsev--Petviashvili equation:
\begin{gather*}
\solution
%u=2\sigma^{3/2}s^{-2}\big(\sqrt3y\big)^{-1}\mathtt h_{\rm IV}'\left(\big(\tfrac\sigma3\big)^{1/4}\frac{x}{s\sqrt y}\right)
%+\sigma\frac{x^2}{24y^4}+2s^{-2}\big(\sqrt3y\big)^{-1}\sigma^{3/2}\lambda.
%\\
u=-9\frac{(3s)^{1/2}}{\sqrt y}\mathtt h_{\rm IV}'\left(\frac{x}{2(s/3)^{1/4}\sqrt y}\right)
-9\lambda\frac{(3s)^{1/2}}{y}+\frac{3x^2}{4y^2}\quad\mbox{with}\quad
s=-\sigma/3,
\end{gather*}

\subsection{Subalgebra family 1.2}\label{sec:RedSub12}

For the subalgebra $\mathfrak s_{1.2}=\langle\mathcal Y(1)=\p_y\rangle$,
it is convenient to use an ansatz $u=v(z_1,z_2)$ with $z_1=t$ and $z_2=x$.
Substituting this ansatz into the equation~\eqref{eq:KP} yields
the derivative with respect to~$z_2$ of the famous Korteweg--de Vries equation:
\begin{equation}\label{eq:KdV1}
(v_1+\tfrac32vv_2+\tfrac14v_{222})_2=0.
\end{equation}
Expanding the left hand side of~\eqref{eq:KdV1} results in
\begin{equation}\label{eq:KdV2}
v_{12}+\tfrac32(v_2)^2+\tfrac32vv_{22}+\tfrac14v_{2222}=0.
\end{equation}

\paragraph{Maximal Lie invariance algebra}
The maximal Lie invariance algebra $\mathfrak g_{1.2}$ of~\eqref{eq:KdV2}
is spanned by the vector fields
\begin{gather}\label{eq:LieInvAlgReduction12}
\p_1,\quad
D^z:=z_1\p_{z_1}+\tfrac13z_2\p_{z_2}-\tfrac23v\p_v,\quad
%G^z:=f(z_1)\p_2+\tfrac23f_{z_1}(z_1)\p_v,
%\\
G^z(f):=f(z_1)\p_2+\tfrac23f_{z_1}\p_v,
\end{gather}
where $f$ is an arbitrary smooth function of $z_1$.
The normalizer ${\rm Nor}_{\mathfrak g}(\mathfrak s_{1.2})$ of the subalgebra $\mathfrak s_{1.2}$
in the algebra $\mathfrak g$ is the span
$\langle\p_t,\p_y, t\p_t+\tfrac13x\p_x+\tfrac23y\p_y-\tfrac23u\p_u,\chi\p_x+\tfrac23\chi_t\p_u\rangle$,
which implies that the Kadomtsev--Petviashvili equation does not have infinitesimal hidden symmetries
under reduction with respect to the subalgebra $\mathfrak s_{1.2}$.

The nontrivial commutation relations between the vector fields~\eqref{eq:LieInvAlgReduction12} are
\[
[\p_{z_1},D^z]=\p_{z_1},\quad 
[D^z,G^z(f)]  =G^z(z_1f_{z_1}-\tfrac13f),\quad
[\p_1,G^z(f)] =G^z(f_{z_1}).
\]
\noprint{
$[D^z,G^z(f)]=z_1f_{z_1}\p_2+\tfrac23z_1f_{z_1z_1}\p_v-\tfrac13f\p_2+\tfrac49f_{z_1}\p_v
=(z_1f_{z_1}-\tfrac13f)+\tfrac23(z_1f_{z_1z_1}+\tfrac23f_{z_1})\p_v
=G^z(z_1f_{z_1}-\tfrac13f)$
}
The vector fields $G^z(f)$ constitute an infinite-dimensional abelian ideal $\mathfrak m_3$ of $\mathfrak g_{1.2}$.
Furthermore, the algebra $\mathfrak g_{1.2}$ splits over $\mathfrak m_3$,
and the corresponding complement subalgebra $\langle\p_{z_1}, D^z\rangle$ is the two-dimensional nonabelian Lie algebra,
$\mathfrak g_{1.2}=\langle\p_{z_1}, D^z\rangle\lsemioplus\mathfrak m_3$.

The algebra $\mathfrak g_{1.2}$ is solvable;
the elements of its derived series are
$\mathfrak g_{1.2}^{(1)}:=[\mathfrak g_{1.2},\mathfrak g_{1.2}]=\langle\p_{z_1}\rangle\lsemioplus\mathfrak m_3$
and $[\mathfrak g_{1.2}^{(1)},\mathfrak g_{1.2}^{(1)}]=\mathfrak m_3$.
This implies that both $\mathfrak m_2:=\mathfrak g_{1.2}^{(1)}$ and $\mathfrak m_3$
are megaideals of $\mathfrak m_1:=\mathfrak g_{1.2}$.
The obtained chain of megaideals 
\[
\mathfrak m_1\varsupsetneq\mathfrak m_2\varsupsetneq\mathfrak m_3
\]
will be instrumental in the course
of computing the point-symmetry pseudogroup of the equation~\eqref{eq:KdV1}.

\begin{remark}\label{rem:RelationWithKdV}
Recall also that the Lie invariance algebra $\mathfrak g^{\rm KdV}$ of the Korteweg--de Vries equation
in the form $v_1+\tfrac32vv_2+\tfrac12v_{222}=0$ is spanned by the vector fields
\[
\p_1,\quad
D^z,\quad
%G^z:=f(z_1)\p_2+\tfrac23f_{z_1}(z_1)\p_v,
%\\
G^z(1)\quad\mbox{and}\quad
G^z(z_1).
\]
Hence, the maximal Lie invariance algebra of the derivative of KdV 
is an infinite-dimensional extension of $\mathfrak g^{\rm KdV}$ by an infinite-dimensional
abelian ideal $\mathfrak m_3:=\langle G(f)\mid f\in C^\infty(\mathbb R)\rangle$.
Note that $\mathfrak m_3\cap\mathfrak g^{\rm KdV}=\langle G^z(1),G^z(z_1)\rangle$,
hence essentially new vector fields that arise as symmetries of the derivative of the KdV
are $G^{z_1}(f)$ with $f_{z_1z_1}\ne0$.
\end{remark}

\paragraph{Point symmetry pseudogroup}

\begin{theorem}\label{thm:PSGKdV_z2}
The point symmetry pseudogroup $G_{1.2}$ of the equation~\eqref{eq:KdV1}
is constituted by the transformations of the form
\begin{gather}\label{eq:PSGKdVDiff}
\tilde z_1=\alpha^3z_1+\lambda,\quad
\tilde z_2=\alpha z_2+h(z_1)\quad
V=\alpha^{-2}v+\tfrac23\alpha^{-3}h_{z_1},
\end{gather}
where $\alpha$ and $\lambda$ are arbitrary real constants with $\alpha\ne0$
and $h(z_1)$ is an arbitrary smooth function of $z_1$.
\end{theorem}

\begin{proof}
To compute the pseudogroup $G_{1.2}$ we apply the megaideal-based version of the algebraic method
(see, e.g., Section~\ref{sec:PSGAlgMethod}) to the above chain of megaideals of $\mathfrak g_{1.2}$.

Consider an arbitrary point transformation $\Phi$ acting on the space
of independent and dependent variables $\mathbb R^2_{z_1,z_2}\times\mathbb R_v$:
\[
\Phi\colon\quad
(\tilde z_1,\tilde z_2,\tilde v)=(Z^1,Z^2,V)(z_1,z_2,v),
\]
with a nonvanishing Jacobian, $|\p(Z^1,Z^2,V)/\p(z_1,z_2,v)|\ne0$.
For $\Phi$ to determine a symmetry of~\eqref{eq:KdV1}
it must preserve the megaideals of the maximal Lie invariance algebra $\mathfrak g_{1.2}$,
in particular, $\Phi_*(\mathfrak m_i)=\mathfrak m_i$ for $i\in\{1,2,3\}$.

Consider the vector field $G^z(f)\in\mathfrak m_3$.
The condition $\Phi_*\mathfrak m_3\subset\mathfrak m_3$ implies that
\[
\Phi_*\big(G^z(f)\big)=G^z(\tilde f^1)
\]
for some smooth function $\tilde f^1(z_1)$.
Composing both sides of this equality with $\Phi$ yields a system of constraints on $Z^1$, $Z^2$ and $V$:
\begin{subequations}
\begin{gather}\label{eq:AlgForKdVM3Cond1}
fZ^1_{z_2}+\tfrac23f_{z_1}Z^1_v=0,
\\\label{eq:AlgForKdVM3Cond2And3}
fZ^2_{z_2}+\tfrac23f_{z_1}Z^2_v=\tilde f^1(\tilde z_1),
\quad
fV_{z_2}+\tfrac23f_{z_1}V_v=\tfrac23\tilde f^1_{\tilde z_1}(\tilde z_1).
\end{gather}
\end{subequations}
Since $f$ is an arbitrary function, splitting \eqref{eq:AlgForKdVM3Cond1} with respect to $f$ and $f_{z_1}$
(which is equivalent to choosing vector fields $G^z(f)$ with $f=1$ or $f=z_1$)
leads to
\begin{gather}\label{eq:KdVPsGConditionsZ1}
Z^1_{z_2}=Z^1_v=0.
\end{gather}
Splitting the compatibility condition between the equations from~\eqref{eq:AlgForKdVM3Cond2And3}
with respect to $f$, $f_{z_1}$ and $f_{z_1z_1}$ yields the constraints
\begin{gather}\label{eq:KdVPsGConditions}
Z^2_v=0,\quad
Z^1_{z_1}V_v=Z^2_{z_2},\quad
Z^1_{z_1}V_{z_2}=\tfrac23Z^2_{z_1z_2}.
\end{gather}
Furthermore, since the right-hand sides of the equations in~\eqref{eq:AlgForKdVM3Cond2And3} 
depend only on $z_1$, it follows that
\begin{gather}\label{eq:KdVPsGCondZ2V}
Z^2_{z_2z_2}=0,\quad
V_{vv}=V_{vz_2}=V_{z_2z_2}=0.
\end{gather}

The pushforward of the vector field $\p_{z_1}\in\mathfrak{m}_2$ takes the form
\[
\Phi_*(\p_{z_1})=\alpha^2\p_{z_1}+G^z(\tilde f^2),
\]
where $\alpha^2$ is a real constant and $\tilde f^2=\tilde f^2(z_1)$ is an arbitrary smooth function of $z_1$.
This leads to a system of constraints on $Z^1$, $Z^2$ and $V$:
\begin{subequations}
\begin{gather} \label{eq:AlgForKdVM2Cond1}
Z^1_{z_1} = \alpha^2, \\ \label{eq:AlgForKdVM2Cond2And3}
Z^2_{z_1} = \tilde f^2(\tilde z_1), \quad
V_{z_1} = \tfrac23 \tilde f^2_{\tilde z_1}(\tilde z_1).
\end{gather}
\end{subequations}
The compatibility condition between the equations from~\eqref{eq:AlgForKdVM2Cond2And3} results in
\begin{gather}\label{eq:KdVPsGCompatibilityForM2}
Z^1_{z_1}V_{z_1}=\tfrac23Z^2_{z_1z_1}.
\end{gather}

Pushing forward the vector field $D^z\in\mathfrak m_1$,
\[
\Phi_*(D^z)=\alpha^3\p_{z_1}+\beta^3D^z+G(\tilde f^3),
\]
where $\alpha^3$, $\beta^3$ are real constants and $\tilde f^3=\tilde f^3(z_1)$ is a smooth function,
yields the following system of constraints on the components of $\Phi$:
\begin{subequations}
\begin{gather} \label{eq:AlgForKdVM1Cond1}
z_1Z^1_{z_1}=\alpha^3+\beta^3Z^1,
\\ \label{eq:AlgForKdVM1Cond2And3}
3z_1Z^2_{z_1}+z_2Z^2_{z_2}-3\beta^3Z^2=3\tilde f(\tilde z_1), \quad
3z_1V_{z_1}+z_2V_{z_2}-2vV_v+2\beta^3V=2\tilde f_{\tilde z_1}(\tilde z_1).
\end{gather}
\end{subequations}
Substituting the second equation from \eqref{eq:KdVPsGConditions} into
the compatibility condition between equations~\eqref{eq:AlgForKdVM1Cond2And3} 
leads to
\begin{gather}\label{eq:KdVPsGConditionOnV}
V=\frac{Z^2_{z_2}}{Z^1_{z_1}}v+\frac{2Z^2_{z_1}}{3Z^1_{z_1}}.
\end{gather}
The general solution of the system constituted by the equations~\eqref{eq:KdVPsGConditionsZ1}--\eqref{eq:KdVPsGCondZ2V},
\eqref{eq:AlgForKdVM2Cond1}, \eqref{eq:KdVPsGCompatibilityForM2} and~\eqref{eq:KdVPsGConditionOnV}
is given by
\begin{gather*}
Z^1=\alpha z_1+\alpha^0,\quad
Z^2=\beta z_2+\beta^0(z_1)\quad
V=\frac{\beta}{\alpha}v+\frac{2\beta^0_{z_1}}{3\alpha}.
\end{gather*}
Using this form for the components of $\Phi$, we proceed by applying the direct method
analogously to Section~\ref{sec:PSGDirectMethod}.
We express the jet monomials in the tilded variables appearing on the left-hand side of~\eqref{eq:KdV2}
in terms of the jet monomials in the original variables:
\begin{gather*}
\tilde v_{\tilde z_1\tilde z_2}=\alpha^{-2}v_{12},\quad
(\tilde v_{\tilde z_2})^2=\alpha^{-2}(v_2)^2,\quad
\tilde v\tilde v_{\tilde z_2\tilde z_2}=\alpha^{-2}vv_{22},\quad
\tilde v_{\tilde z_2\tilde z_2\tilde z_2\tilde z_2}=\alpha^{-1}\beta^{-3}v_{2222}.
\end{gather*}
Therefore, the transformed equation $\Phi_*$\eqref{eq:KdV2} is satisfied for all solutions of~\eqref{eq:KdV2} 
if and only if $\alpha=\beta^3$.
Relabeling $(\beta,\beta^0(z_1),\alpha^0)\mapsto(\alpha,h(z_1),\lambda)$ results in theorem's statement.
\end{proof}

The generalized Galilean boosts
\[
\tilde z_1=z_1,\quad
\tilde z_2=z_2+h(z_1)\quad
V=v+\tfrac23\alpha^{-3}h_{z_1}
\]
constitute an infinite-dimensional normal (commutative) pseudosubgroup of $G_{1.2}$,
which we denote by $\mathcal B^z$ (boosts).
The pseudogroup $G_{1.2}$ splits over $\mathcal B^z$, $G_{1.2}=G_{1.2}^{\rm aff}\ltimes\mathcal B^z$,
where the group $G_{1.2}^{\rm aff}$ is the subgroup of $G_{1.2}$ constituted by the transformations
of the form~\eqref{eq:PSGKdVDiff} with $h=0$.
The group $G_{1.2}^{\rm aff}$ is isomorphic to the rank-one affine group ${\rm Aff}(1,\mathbb R)$,
or, equivalently, the semidirect product $\mathbb R^\times\ltimes\mathbb R$,
where the multiplication is given by $(\alpha_0,\beta_0)\cdot(\alpha,\beta)=(\alpha_0\alpha,\alpha_0\beta+\beta_0)$.
Therefore, the identity component of $G_{1.2}$ is constituted by the transformations~\eqref{eq:PSGKdVDiff} with $\alpha>0$,
hence a complete list of discrete symmetries of the equation~\eqref{eq:KdV2}
that is independent up to composing them with each other and with elements of the identity component of  $G_{1.2}$
are exhausted by the involutions of simultaneous alternating signs of $z_1$ and $z_2$,
$(z_1,z_2,v)\mapsto(-z_1,-z_2,v)$.
The quotient of the pseudogroup $G_{1.2}$ by its identity component
is isomorphic to~$\mathbb Z_2$.

A necessary condition for a point symmetry $\Phi$ of the Kadomtsev--Petviashvili equation~\eqref{eq:KP}
to induce a point symmetry of the reduced equation~\eqref{eq:KdV1}
is the preservation of the foliation of the space $\mathbb R^4_{t,x,y,u}$ 
associated with the vector field $\p_y$.
That is, $\Phi$ preserves the foliation cut out by the equation $t=c_1$, $x=c_2$ and $u=c_3$,
where $c_1$, $c_2$ and $c_3$ are arbitrary real constants.
Therefore, the components of $\Phi$ given in~\eqref{eq:PointSymGroup} additionally satisfy the differential equations
\[
T_{tt}=0\quad\mbox{and}\quad Y^0_t=0,
\] 
which result in the transformations of the form
\[
\tilde t=\tau_1 t+\tau_0,\quad
\tilde x=\tau_1^{1/3}x+X^0(t),\quad
\tilde y=\varepsilon\tau_1^{2/3}y+y_0,\quad
\tilde u=\tau_1^{-2/3}u+\tfrac23\tau_1^{-1}X^0_t,
\]
where $\tau_1$ and $y_0$ are arbitrary real constants with $\tau^1\ne0$,
$\varepsilon=\pm1$
and $X^0(t)$ is an arbitrary smooth function of $t$.
Hence, all point symmetries of~\eqref{eq:KdV1} are entirely induced
from the point symmetries of the  Kadomtsev--Petviashvili equation~\eqref{eq:KP}.

\paragraph{Lie reductions}

Lie reductions and exact invariant solutions of the Korteweg--de Vries equation are well known
and, for instance, can be found in~\cite{bihl2015a}.
As we shall see in what follows, the essential Lie reductions for the derivative of the KdV equation
arise from the reductions of the KdV equation itself;
thus, the presence of the derivative does not play a significant role.
Nevertheless, for the sake of completeness, we present an analysis of the Lie reductions for equation~\eqref{eq:KdV1}.

Following the general procedure of Lie reductions, see, e.g., \cite[Section~2]{vinn2024a},
we begin with classification of one-dimensional subalgebras of $\mathfrak g_{1.2}$.

\begin{proposition}\label{prop:SubalgKdV}
A complete list of $G_{1.2}$-inequivalent subalgebras of the maximal Lie invariance algebra $\mathfrak g_{1.2}$
the equation \eqref{eq:KdV1} is exhausted by the following subalgebras:
\[
\langle D^z \rangle,\quad
\langle\partial_1+G^z(f)\rangle,\quad
\langle G^z(f)\rangle,
\]
where $f(z_1)$ is an arbitrary smooth function.
\end{proposition}
\begin{proof}
Consider the most general form of the vector field $Q$ that belongs to the algebra $\mathfrak g$,
\[
Q:=c_1\partial_1+c_2 D^z+G^z(f).
\]
The pushforward of $Q$ by a point symmetry $\Phi\in G_{1.2}$~\eqref{eq:PSGKdVDiff}
takes the following form:
\[
\Phi_{*}(Q)=(\alpha^3c_1-\lambda c_2)\partial_1+c_2 D^z
+D^z\big(f(\alpha^{-3}(z_1 - \lambda))-\tfrac{c_2}3h(\alpha^{-3}(z_1 - \lambda))+c_2z_1h_{z_1}(\alpha^{-3}(z_1 - \lambda))\big).
\]

If $c_2 \ne 0$, then setting the parameters $\alpha=1$, $\lambda=c_1/c_2$ and
choosing $h$ to be satisfy the ordinary differential equation $z_1h_{z_1}-\tfrac13 h+c_2^{-1}f = 0$,
we therefore obtain subalgebra $\langle D^z\rangle$.

If $c_2=0$ and $c_1\ne0$, that is $Q=c_1\partial_1+G^z(f)$, then
\[
\Phi_*(Q)=\alpha^3c_1\partial_1+G^z\big(f(\alpha^{-3}(z_1 - \lambda))\big).
\]
By this we obtain the subalgebra family $\langle\partial_1+G^z(f)\rangle$,
parameterized by an arbitrary smooth function $f(z_1)$.

For the last case of $c_1=c_2=0$, the arguments are similar to the previous case.
This results in the family of subalgebras $\langle G^z(f)\rangle$
\end{proof}

For each subalgebra $\mathfrak s$ listed in Proposition~\ref{prop:SubalgKdV},
we write a suitable $\mathfrak s$-invariant ansatz and the corresponding reduced equation. 
We arrange them into Table~\ref{tab:LieRedKdV1}.

\begin{table}[!ht]\footnotesize
\begin{center}
\caption{$G_{1.2}$-inequivalent codimension-one Lie reductions of the KdV consequence \eqref{eq:KdV1}}
\label{tab:LieRedKdV1}
${ }$\\[-1ex]
\renewcommand{\arraystretch}{1.6}
\begin{tabular}{|l|l|c|c|l|l|}
\hline 
no. & \!Subalgebra\!                        & $v$              			 & $\omega$    			   			& \hfil Reduced equation \\
\hline
1. & $\langle D^z \rangle$                  &$z_1^{-2/3}\varphi$         		  & $z_2 / z_1^{1/3}$       		 & $(\varphi_{\omega\omega\omega}+6\varphi \varphi_{\omega} - \tfrac43 \omega \varphi_{\omega} - \tfrac83 \varphi)_{\omega} = 0$\\
2. & $\langle\partial_1 + G^z(f_{z_1})\rangle$    &$\varphi + \tfrac23 f_{z_1}(z_1)$  		  & $z_2-f(z_1)$   & 			$(\varphi_{\omega\omega\omega} + 6 \varphi \varphi_{\omega})_{\omega} = 0$ \\
3. & $\langle G^z(f) \rangle$               &$\varphi + \tfrac23 z_2 f_{z_1}/f(z_1)$ & $z_1$   						 &  	$f_{z_1 z_1} = 0$ \\
\hline
\end{tabular}\\[2ex]
Here $\varphi=\varphi(\omega)$ is the new unknown function of the invariant independent variable~$\omega$.
\end{center}
\end{table}

As is evident from Table~\ref{tab:LieRedKdV1}, the reduction with respect to the subalgebra $\langle G^z(f) \rangle$ 
leads to a differential constraint $f_{z_1z_1}=0$ on the subalgebra parameter.
The vector fields that satisfy these constraints are symmetries of the standard Korteweg--de Vries equation,
as noticed in Remark~\ref{rem:RelationWithKdV}.
Since vector fields $G(f)$ with $f_{z_1z_1}\ne0$ were essentially new symmetries arising in~\eqref{eq:KdV1}
compared with the standard KdV, Lie reductions of~\eqref{eq:KdV1} are essentially those of KdV.
Nevertheless, we will integrate them and pull back the obtained solutions
to the solutions of the Kadomtsev--Petviashvili equation~\eqref{eq:KP}.

\medskip\par\noindent
{\bf Reduction 1.}
Integrating once the reduced equation leads to
\begin{gather*}\label{eq:Reduction1IntKdV}
\varphi_{\omega\omega\omega}+6\varphi \varphi_{\omega}-\tfrac43\omega\varphi_{\omega}-\tfrac83\varphi=c.
\end{gather*}
Using the Miura transform
$\varphi=\tfrac{a^2}6(\psi_{\tilde\omega}-\tfrac16\psi^2)$,
where $\psi=\psi(\tilde\omega)$ with $\tilde\omega=a\omega$, $a:=(4)^{1/3}$,
we obtain the equation
\begin{gather}\label{eq:Miura}
W_{\tilde\omega\tilde\omega}-\frac13\psi W_{\tilde\omega}=\frac{3c}{2a^2},
\end{gather}
where 
\[
W(\tilde\omega):=\psi_{\tilde\omega\tilde\omega}-\frac1{18}\psi^3-\frac13\tilde\omega\psi-k.
\]
In the particular case of $c=0$, an obvious solution of~\eqref{eq:Miura} is $W(\tilde\omega)=0$,
which leads to
\[
\psi_{\tilde\omega\tilde\omega}=\frac1{18}\psi^3+\frac13\tilde\omega\psi+k.
\]
The latter equation is reduced to Painlev\'e II via the change of variables $\psi=2\hat\psi(\hat\omega)$
and $\hat\omega=3\tilde\omega$:
\[
\hat\psi_{\hat\omega\hat\omega}=2\hat\psi^3+\hat\omega\hat\psi+\alpha,
\]
where $\alpha$ is an arbitrary real constant.
Hence, taking a solution $\mathtt p_{\rm II}(\hat\omega)$ of Painlev\'e II,
we pull it back to the solution of the Korteweg--de Vries equation~\eqref{eq:KdV1},
\begin{gather*}
\solutionRedEq
v=\frac{a^2}3z_1^{-2/3}\Big(\mathtt p_{\rm II}'\big(3az_2z_1^{-1/3}\big)-\tfrac13\mathtt p_{\rm II}\big(3az_2z_1^{-1/3}\big)^2\Big)
\end{gather*}
and thus to the solution of the Kadomtsev--Petviashvili equation~\eqref{eq:KP}:
\[
\solution
u=\frac{a^2}3t^{-2/3}\Big(\mathtt p_{\rm II}'\big(3axt^{-1/3}\big)-\tfrac13\mathtt p_{\rm II}\big(3axt^{-1/3}\big)^2\Big).
\]
Here $a:=4^{1/3}$ and $\mathtt p_{\rm II}'(\omega)$ denotes the derivative of $\mathtt p_{\rm II}(\omega)$ with respect to $\omega$.

\medskip\par\noindent
{\bf Reduction~2}.
The reduced equation can be easily integrated twice:
\[
\varphi_{\omega\omega}+3\varphi^2=c_1\omega+c_0.
\]
Using substitution $\varphi(\omega)=-2a^2\tilde\varphi(\tilde\omega)$ with $\tilde\omega=a\omega-c_0/(2a^4)$, $a^5:=-c_1/2$,
we derive Painlev\'e~I equation:
\[
\tilde\varphi_{\tilde\omega\tilde\omega}=6\tilde\varphi^2+\tilde\omega.
\]
The pullback of a solution $\tilde\varphi=\mathtt{p}_{\rm I}$ of the Painlev\'e I
leads to the solution of the Korteweg--de Vries equation
\begin{gather*}
\solutionRedEq
v=-2a^2\mathtt{p}_{\rm I}\big(a(z_2-f(z_1))-c_0/(2a^4)\big)+\frac23f_{z_1}(z_1)
\end{gather*}
and, consequently, to the solution of the Kadomtsev--Petviashvili equation:
\begin{gather*}
\solution
u=-2a^2\mathtt{p}_{\rm I}\big(a(x-f(t))-c_0/(2a^4)\big)+\frac23f_{t}(t),
\end{gather*}
where $a$ is an arbitrary real constant.

\medskip\par\noindent
{\bf Reduction 3.}
This is a trivial reduction, which merely leads to the following solution of Kadomtsev--Petviashvili equation
\[
\solution
u=h(t),
\]
where $h$ is an arbitrary smooth function of $t$.

\subsection{Subalgebra family 1.3}

The subalgebra $\mathfrak s_{1.3}=\langle\p_x\rangle$ leads to an invariant ansatz
$u=v(z_1,z_2)$ with $z_1=y$, $z_2=t$,
which reduces the equation~\eqref{eq:KP} to a homogeneous
second-order linear ordinary differential equation with respect to the variable $z_1$:
\begin{gather}\label{eq:KPReduction13}
v_{11}=0,
\end{gather}
and $z_2$ plays the role of a parameter.
The general solution of~\eqref{eq:KPReduction13} is a linear function of $z_1$:
\[
v=c^0(z_2)z_1+c^1(z_2),
\]
where $c^0$ and $c^1$ are arbitrary smooth functions of $z_2$.
Consequently, the corresponding exact solution of the Kadomtsev--Petviashvili equation is
\begin{gather}
\solution
u=c^0(t)y+c^1(t),
\end{gather}
which, expectedly, lacks physical relevance.

For this reason, reduction with respect to $\mathfrak s_{1.3}$
is scarcely considered in the literature in the context of the Kadomtsev--Petviashvili equation~\eqref{eq:KP}.
However, the equation~\eqref{eq:KPReduction13} admits a large Lie symmetry algebra,
which yields families of hidden symmetries for~\eqref{eq:KP}.

The maximal Lie invariance algebra $\mathfrak g_{1.3}$ of the equation \eqref{eq:KPReduction13} is the span of
\begin{gather*}
\tau(z_2)\p_{z_2},\quad
c^0(z_2)\p_{z_1},\quad
c^1(z_2)z_1\p_{z_1},\quad
c^2(z_2)\left(z_1^2\p_{z_1}+z_1v\p_v\right),\quad
f(z_2)v\p_{z_1},
\\
d^0(z_2)\p_v,\quad
d^1(z_2)v\p_v,\quad
d^2(z_2)\left(z_1v\p_{z_1}+v^2\p_v\right),\quad
g(z_2)z_1\p_v,
\end{gather*}
where $\tau$, $c^i$, $d^i$, $f$, $g$, $i\in\{0,1,2\}$ are arbitrary smooth functions depending on $z_2$.
Viewed as a module over the ring of smooth functions $C^{\infty}(\mathbb R_{z_2})$,
the algebra is generated by the vector fields
\begin{gather*}
\p_{z_2},\quad
\p_{z_1},\quad
z_1\p_{z_1},\quad
z_1^2\p_{z_1}+z_1v\p_v,\quad
v\p_{z_1},\quad
\p_v,\quad
v\p_v,\quad
z_1v\p_{z_1}+v^2\p_v,\quad
z_1\p_v.
\end{gather*}
Notice that the subalgebra $\langle\p_{z_1},
z_1\p_{z_1},
z_1^2\p_{z_1}+z_1v\p_v,
v\p_{z_1},
\p_v,
v\p_v,
z_1v\p_{z_1}+v^2\p_v, 
z_1\p_v\rangle$
is isomorphic to the Lie algebra ${\rm sl}(3,\mathbb R)$.

The normalizer ${\rm Nor}_{\mathfrak g}(\mathfrak s_{1.3})$ of the subalgebra $\mathfrak s_{1.3}$
in the algebra $\mathfrak g$ consists of the elements of the form
$\mathcal X(\chi) + \mathcal Y(\rho) + \mathcal T(\tau_1 t + \tau_0)$.
The algebra of induced infinitesimal symmetries is spanned by the vector fields
\[
\p_{z_2},\quad
z_2\p_{z_2}+\tfrac23z_1\p_{z_1}-\tfrac23v\p_v,\quad
f(z_2)\p_v,\quad
\tfrac{3\sigma}2h(z_2)\p_{z_1}+\tfrac23z_1h_{z_2z_2}\p_v.
\]

\begin{theorem} \label{thm:PSG3}
The point symmetry pseudogroup $G_{1.3}$ of the equation~\eqref{eq:KPReduction13}
consists of transformations of the form
\begin{gather*}
\tilde z_1=\frac{a^1(z_2)z_1+b^1(z_2)v+c^1(z_2)}{A(z_2)z_1+B(z_2)v+C(z_2)},\quad
\tilde z_2=Z^2(z_2),\quad
\tilde v  =\frac{a^2(z_2)z_1+b^2(z_2)v+c^2(z_2)}{A(z_2)z_1+B(z_2)v+C(z_2)},
\end{gather*}
where $a_1,b_1,c_1,a_2,b_2,c_2,A,B,C, Z^2$ are arbitrary smooth functions of $z_2$
with
\begin{gather*}
Z^2_{z_2}\ne0\quad\mbox{and}\quad
\begin{vmatrix}
a^1(z_2) & b^1(z_2) & c^1(z_2)\\
a^2(z_2) & b^2(z_2) & c^2(z_2)\\
A(z_2)   & B(z_2)   & C(z_2)
\end{vmatrix}\ne0.
\end{gather*}
\end{theorem}

\begin{proof}
To find the pseudogroup $G_{1.3}$, we use the direct method, see Section~\ref{sec:PSGDirectMethod}.
Consider the most general form of a point transformation
acting on the space $\mathbb R^3_{z_1,z_2,v}$,
\begin{gather*}
\Phi\colon\quad
(\tilde z_1,\tilde z_2,\tilde v)=
(Z^1,Z^2,V)(z_1,z_2,v),
\end{gather*}
with the nondegenerate Jacobian.
Using chain rule, we express $v_{z_1z_1}$ in terms of derivatives of $\tilde v$ with respect to $\tilde z_1$ and $\tilde z_2$,
and substitute $\tilde v_{\tilde z_2\tilde z_2}=0$.
For these, we act at $\tilde v=V(z_1,z_2,v)$ by the operators of total derivatives ${\rm D}_{z_1}$ and ${\rm D}_{z_1}^2$.

Splitting with respect to the jet variable $\tilde v_{\tilde z_2\tilde z_2}$ leads to 
\begin{gather}\label{eq:Red3PSGCond}
D_{z_1}Z^2=0,\quad
D_{z_1}^2V-(D_{z_1}^2 Z^1)\tilde v_{\tilde z_1}=0
\end{gather}
The first equation from~\eqref{eq:Red3PSGCond} straightforwardly leads to $Z^2_{z_1}=0$.
Splitting the second equation from~\eqref{eq:Red3PSGCond} with respect to
$v_{z_1}^3$, $v_{z_1}^2$, $v_{z_1}$ and $1$ results in the following system
of partial differential equations with respect to $z_1$ and $v$
\begin{gather*}
Z^1_vV_{vv}-V_v Z^1_{vv}=0,
\\
Z^1_{z_1}V_{vv}-V_{z_1}Z^1_{vv}+2(Z^1_vV_{z_1 v}-V_vZ^1_{z_1 v})=0,
\\
Z^1_vV_{z_1 z_1}-V_v Z^1_{z_1 z_1}+2(Z^1_{z_1}V_{z_1 v}-V_{z_1}Z^1_{z_1 v})=0,
\\
Z^1_{z_1} V_{z_1 z_1}- V_{z_1} Z^1_{z_1 z_1}=0,
\end{gather*}
where $z_2$ plays role of a parameter.
Such a system is known in the literature as a two-dimensional Schwarzian system,
see, e.g., \cite[(29)]{piza2016a}.
Its general solution is given by linear fractional transformations in the variables $(z_1,v)$,
where the coefficients depend arbitrarily on $z_2$.
No further constraints are imposed on the components of $\Phi$.
\end{proof}

A necessary condition for a point symmetry $\Phi$ of the Kadomtsev--Petviashvili equation~\eqref{eq:KP}
to induce a point symmetry of the reduced equation~\eqref{eq:KPReduction13}
is the preservation of the foliation of the space $\mathbb R^4_{t,x,y,u}$ 
associated with the vector field $\p_x$.
That is, $\Phi$ preserves the foliation cut out by the equation $t=c_1$, $y=c_2$ and $u=c_3$,
where $c_1$, $c_2$ and $c_3$ are arbitrary real constants.
Therefore, the components of $\Phi$ given in~\eqref{eq:PointSymGroup} additionally satisfy the differential equation
\[
T_{tt}=0,
\] 
which result in the transformations of the form
\begin{align*}
\tilde t&=\tau_1 t+\tau_0,\quad
\tilde x=\tau_1^{1/3}x+ \frac{2 \sigma \epsilon Y^0_t}{3 \tau_1^{1/3}}y +X^0(t),\quad
\tilde y=\varepsilon\tau_1^{2/3}y+Y^0(t),\\
\tilde u&=\tau_1^{-2/3}u + \frac{4 \sigma \epsilon Y^0_{tt}}{9 \tau_1^{4/3}}y + \frac{2 X^0_t}{3\tau_1} - 
\frac{2 \sigma (Y^0_t)^2}{9 \tau_1^2},
\end{align*}
where $\tau_1$ and $\tau_0$ are arbitrary real constants with $\tau^1\ne0$,
$\varepsilon=\pm1$
and $X^0(t), Y^0(t)$ are arbitrary smooth functions of $t$.

Then, the induced transformations have the form
\[
\tilde z_1 = \varepsilon \tau_1^{2/3} z_1 + c^1(z_2), \quad
\tilde z_2 = \tau_1 z_2 + \tau_0, \quad
\tilde v = \tau_1^{-2/3} v + \frac{4 \sigma \epsilon }{9 \tau_1^{4/3}} c^1_{z_2 z_2} z_1 + c^2(z_2),
\]
where $\tau_1$ and $\tau_0$ are arbitrary real constants with $\tau^1\ne0$,
$\varepsilon=\pm1$
and $c_1(z_2), c_2(z_2)$ are arbitrary smooth functions of $z_2$.
Obviously, every such transformation is indeed an induced symmetry of the equation \eqref{eq:KPReduction13}.
\section{Codimension-two reductions}\label{sec:Codim2Reductions}

\begin{table}[!ht]\footnotesize
\begin{center}
\caption{$G$-inequivalent codimension-two Lie reductions of the KP equation \eqref{eq:KP} }
\label{tab:LieRedKP2}
${ }$\\[-1ex]
\renewcommand{\arraystretch}{1.6}
\begin{tabular}{|l|l|c|c|l|l|}
\hline 
no. & \!Subalgebra\!                        	& $u$              	          & $\omega$    			    & \hfil Reduced equation \\
\hline
1. & $\langle\partial_t,\partial_y\rangle$		& $\varphi$         			 & $x$       		 	& $(\varphi_{\omega\omega\omega}+3(\varphi^2)_\omega)_\omega=0$\\
2. & $\langle\partial_t,\partial_x\rangle$    	& $\varphi$  		  			 & $y$   				& $\varphi_{\omega\omega} = 0$ \\
3. & $\langle\partial_y,h(t)\partial_x+\tfrac23h_t\partial_u\rangle$ 
                                                & $(\varphi+\tfrac23h_tx)$ 	 & $t$   				& $h_{tt}=0$ \\
4. & $\langle\partial_t,\mathcal T(t)\rangle$   & $\tfrac{\sigma\sqrt3}{6y}(\varphi+\tfrac14\tfrac{x^2\sqrt3}y)$ 	 			 & $3^{\tfrac14}x|y|^{-\tfrac12}$   &  	
$\sigma\varphi_{\omega\omega\omega\omega}+\tfrac12\epsilon(\varphi^2)_{\omega\omega}=\tfrac34\big((\omega\varphi)_\omega+\varphi\big)+\tfrac98\epsilon\omega^2$ \\
5. & $\langle\partial_y,\mathcal T(t)\rangle$   & $t^{-\tfrac23}\varphi$ 	 & $x t^{-\tfrac13}$   	
 & $(\varphi_{\omega\omega\omega}+3(\varphi^2)_\omega-\tfrac43(\omega\varphi_\omega+2\varphi))_\omega=0$ \\
6. & $\langle\partial_x,\mathcal T(t)\rangle$   & $y^{-1}\varphi$ 	 	 		 & $y t^{-\tfrac23}$   	&  $\omega^2\varphi_{\omega\omega}-2\omega\varphi_\omega+2\varphi=0$ \\
\hline
\end{tabular}\\[2ex]
Here $\varphi=\varphi(\omega)$ is the new unknown function of the invariant independent variable~$\omega$,
$\epsilon:=\sgn y$.
\end{center}
\end{table}

In fact, all codimension-two Lie reductions of the Kadomtsev--Petviashvili equation~\eqref{eq:KP} are superfluous.
We have the following inclusions of subalgebras:
\[
\mathfrak s_{1.1}\subset\mathfrak s_{1.2},\quad
\mathfrak s_{1.3}\subset\mathfrak s_{2.2},\quad
\mathfrak s_{1.2}\subset\mathfrak s_{2.3}^h,\quad
\mathfrak s_{1.1}\subset\mathfrak s_{2.5},\quad
\mathfrak s_{1.2}\subset\mathfrak s_{2.6},\quad
\mathfrak s_{1.3}\subset\mathfrak s_{2.7}.
\]
and the subalgebra $\mathfrak s_{2.4}^f$ does not satisfy the local transversality condition,
hence not appropriate for constructing ansatz.
Furthermore, for any inclusion $\mathfrak s_1\subset\mathfrak s_2$ listed above,
the complement vector field to the subalgebra $\mathfrak s_1$ in $\mathfrak s_2$
induces a Lie symmetry of the reduced model.
This fact guarantees that any solution that is invariant with respect to a two-dimensional subalgebra
is essentially a two-step reduction of~\eqref{eq:KP}, where the first step is a one-dimensional subalgebra
listed in Theorem~\ref{thm:1DSubalgebrasKP} and the second step is a subalgebra
of the maximal Lie invariance algebra of the obtained reduced model.

Nevertheless, to each appropriate two-dimensional subalgebra listed in Theorem~\ref{thm:2DSubalgebrasKP},
we associate an invariant ansatz and derive the corresponding reduced system in Table~\ref{tab:LieRedKP2}.

The integration of reduction 1 and reduction 5 from Table~\ref{tab:LieRedKP2}
was performed in cases Reduction 2 and Reduction 1 of Section~\ref{sec:RedSub12}, respectively.
Reduction~4 is precisely Reduction~3 arising in Section~\ref{sec:Codim1Reductions}.
Finally, Reduction~6 easily integrates to
\begin{gather*}
\varphi=c_1\omega^2+c_2\omega,
\end{gather*}
leading to the following solution of the Kadomtsev--Petviashvili equation:
\begin{gather*}
u=c_1 yt^{-\tfrac43}+c_2t^{-\tfrac23}.
\end{gather*}

\section{Conclusion}\label{sec:Conclusion}

The Kadomtsev--Petviashvili equation~\eqref{eq:KP} is one of the most well-studied integrable system in 1+2 dimensions.
Surprisingly, a nice and closed-form description of its point symmetries has been missing from the literature,
although given how complex the underlying computations are, that is completely understandable.
Previous works were typically focused on exponentiating the maximal Lie invariance algebra $\mathfrak g$ of~\eqref{eq:KP}, see, e.g., \cite{davi1986a},
supplemented by discrete point symmetries found through heuristic considerations.
The main goal of this paper is to fill that gap by determining the complete point symmetry pseudogroup of the KP equation,
classifying its subalgebras, carrying out Lie reductions and constructing the corresponding invariant solutions.

We determined the complete point symmetry pseudogroup~$G$ of the Kadomtsev--Petviashvili equation~\eqref{eq:KP} in Theorem~\ref{thm:PointSymGroupKP}
using both direct and algebraic methods, thereby independently verifying the obtained results
and illustrating the applicability of these approaches.
While the algebraic method is known to be a robust technique that can be applied relatively straightforwardly
to a single system of differential equations (either linear or nonlinear), see, e.g., \cite{kova2026c,kova2025a,boyk2024a},
the direct method can rarely be applied to a single system without applying additional tricks,
such as embedding the system into a class of systems of differential equations (even in the linear case), see, e.g., \cite{kova2023a,kova2023b}.
Therefore, it is worth noting that we were able to perform all computations in Section~\ref{sec:PSGDirectMethod}
exclusively using the direct method.

The identity component of $G$ and the complete set of its discrete symmetries were established in Corollary~\ref{eq:DiscreteSymmetriesOfKP}.
We also rigorously computed the pseudogroup of contact symmetries $G_c$ of~\eqref{eq:KP},
and established that it is the first prolongation of the pseudogroup $G$, $G_c=G_{(1)}$.

The classification of one- and two-dimensional subalgebras of the maximal Lie invariance algebra $\mathfrak g$
is an essential prerequisite for constructing inequivalent Lie reductions of the KP equation.
In contrast to the conventional classification modulo the group of inner automorphisms of $\mathfrak g$,
we perform this classification modulo the complete point symmetry pseudogroup~$G$
in Theorems \ref{thm:1DSubalgebrasKP} and \ref{thm:2DSubalgebrasKP}.
This approach additionally takes into account the discrete symmetries of the equation 
and provides a systematic basis for classifying $G$-inequivalent reductions.

The codimension-one Lie reductions of the KP equation have led to the Boussinesq equation \eqref{eq:Red11Boussinesq},
the differentiated Korteweg--de Vries equation \eqref{eq:KdV1},
and the homogeneous second-order linear ordinary differential equation \eqref{eq:KPReduction13}.
The complete point symmetry pseudogroups of these reduced equations were determined in Theorems \ref{thm:PSGBoussinesq}, \ref{thm:PSGKdV_z2}, 
and \ref{thm:PSG3}, respectively.
Furthermore, the induced symmetries were identified, and it was shown that all point symmetries of the reduced Boussinesq
and differentiated KdV equations are induced by symmetries of the original KP equation.
The corresponding invariant solutions were constructed explicitly or expressed in terms of solutions of Painlev\'e equations.
In contrast to previous works on this subject, e.g., \cite{bihl2015a, clar1989a, davi1986a},
we presented all substitutions that reduced the reduced models to integrable ordinary differential equations.
The obtained solutions were then pulled back to the solutions of Kadomtsev--Petviashvili equation~\eqref{eq:KP}.

We also constructed and analyzed the codimension-two Lie reductions.
While the resulting equations yielded no essentially new invariant solutions,
we associated an invariant ansatz and a reduced ordinary differential equation with each suitable two-dimensional subalgebra.
The derived reduced equations were then associated with the two-step reductions performed in Section~\ref{sec:Codim1Reductions}.

This work is part of an ongoing program investigating the Lie and point symmetries of integrable systems \cite{boyk2024a,kova2023a,malt2024a,vinn2024a,vinn2026a}.
A natural next step is to examine the symmetries and exact invariant solutions of different representations of the Kadomtsev–Petviashvili equation,
such as its Hirota bilinear form and Lax representation.
Furthermore, building on previous studies concerning the nonclassical symmetries and Lie reductions of~\eqref{eq:KP}, \cite{clar1991a},
extending these results to the bilinear and Lax representations seems to be a promising avenue for future investigation.

\section*{Acknowledgments}
%\medskip\par\noindent{\bf Acknowledgements.}

\begin{table}[!ht]
\begin{minipage}{20mm}
    \includegraphics[width=20mm]{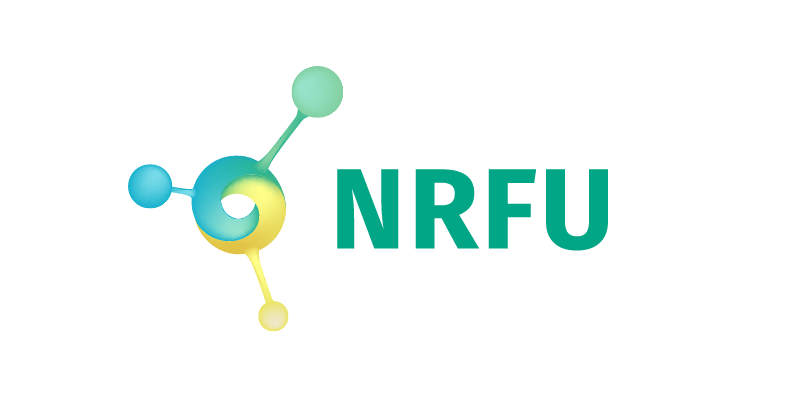}
\end{minipage}
\
\begin{minipage}{138mm}%}
Y.Y.C. expresses gratitude to the National Research Foundation of Ukraine, thanks to whose grant support the project 2025.07/0405 ``Algebraic methods for studying equations of mathematical physics'' is being implemented.
The contents of this work do not necessarily reflect the views of the National Research Foundation of Ukraine and are the sole responsibility of the Institute of Mathematics of NAS of Ukraine.\looseness=-1
\end{minipage}
\end{table}
\vspace{-4.2mm}
\noindent
The research of S.D.K. was supported by the AARMS Postdoctoral Fellowship.
The authors are grateful to Roman O. Popovych for valuable discussions.

\end{document}